\documentclass[apj]{emulateapj}
\usepackage{apjfonts}
\usepackage{times}
\usepackage{amsmath}
\usepackage{graphicx}
\usepackage{natbib}
\slugcomment{Accepted for publication in ApJS}
\shorttitle{Revisiting the Virial Factor}
\shortauthors{Park et al.}

\newcommand{\mbh}{$M_{\rm BH}$}
\newcommand{\msun}{$M_{\odot}$}
\newcommand{\msigma}{$M_{\rm BH}-\sigma_{*}$}
\newcommand{\kms}{km~s$^{\rm -1}$}

\begin{document}

\title{Recalibration of the virial factor and \msigma~relation for local active galaxies}
\author{Daeseong Park$^{1}$}
\author{Brandon C. Kelly$^{2}$}
\author{Jong-Hak Woo$^{1,\dagger}$}
\author{Tommaso Treu$^{2}$}
\affil{$^{1}$Astronomy Program, Department of Physics and Astronomy, Seoul National University, Seoul, 151-742, Republic of Korea; pds2001@astro.snu.ac.kr, woo@astro.snu.ac.kr}
\affil{$^{2}$Department of Physics, University of California, Santa Barbara, CA 93106, USA; bckelly@physics.ucsb.edu, tt@physics.ucsb.edu}
%
\altaffiltext{$\dagger$}{Author to whom correspondence should be addressed.}

\begin{abstract}
Determining the virial factor of the broad-line region (BLR) gas is crucial for calibrating AGN black hole mass
estimators, since the measured line-of-sight velocity needs to be converted into the intrinsic virial velocity.
The average virial factor has been empirically calibrated based on the \msigma~relation of quiescent galaxies,
but the claimed values differ by a factor of two in recent studies. We investigate the origin of the difference
by measuring the \msigma~relation using an updated galaxy sample from the literature, and explore the
dependence of the virial factor on various fitting methods. We find
that the discrepancy is primarily caused by the sample selection, 
while the difference stemming from the various regression methods is marginal.
However, we generally prefer the \texttt{FITEXY} and \texttt{Bayesian} 
estimators based on Monte Carlo simulations for the \msigma~relation. 
In addition, the choice of independent variable in the regression
leads to $\sim 0.2$ dex variation in the virial factor inferred from the calibration process.
Based on the determined virial factor, we present the updated \msigma~relation of local active galaxies.
\end{abstract}
\keywords{galaxies: quiescent -- galaxies: active -- scaling relation -- virial factor -- regression methods}

\section{INTRODUCTION}

Supermassive black holes (SMBHs) are thought to be ubiquitous in the centers of virtually all 
massive galaxies (e.g., Kormendy \& Richstone 1995; Richstone et al. 1998; Ferrarese \& Ford 2005). 
The close connection of black hole growth to galaxy evolution is inferred from the discovery of tight 
correlations between the masses of SMBHs (\mbh) and the global properties of host galaxies, such as 
the stellar velocity dispersion ($\sigma_*$; Ferraresse \& Merritt 2000; Gebhardt et al. 2000)
bulge luminosity ($L_{\rm bul}$; Magorrian et al. 1998; Marconi \& Hunt 2003),
and bulge mass ($M_{\rm bul}$; H\"aring \& Rix 2004).
The origin of these connections has been investigated in theoretical studies of galaxy evolution either through the introduction of active galactic nucleus (AGN) feedback (e.g., Kauffmann \& Haehnelt 2000; Di Matteo et al. 2005; Croton et al. 2006; Hopkins et al. 2006; Bower et al. 2006; Somerville et al. 2008; Booth \& Schaye 2009) or as simply being the result of a hierarchical merging framework (e.g., Peng 2007; Hirschmann et al. 2010; Janke \& Maccio 2011).
The interplay between the black holes and galaxies is now one of the basic ingredients 
in our understanding of galaxy formation and evolution.

In order to better understand the origin and evolution of the SMBH-host galaxy connection, AGN demographics, 
and the growth of the SMBHs through cosmic time, 
an accurate and precise measurement of black hole mass is essential.
Stellar/gas dynamical modeling is commonly used to measure the black hole masses in quiescent galaxies.
However, this technique requires high spatial resolution to resolve the sphere-of-influence 
of the black hole, thereby limiting it to the local universe.
In active galaxies, the black hole mass can be determined by utilizing AGN variability.
The reverberation mapping technique (Blandford \& McKee 1982; Peterson 1993) has been used to estimate the mean size of the broad line region (BLR, $R_{\rm BLR}$) by cross-correlating the continuum light curve with the broad emission line light curve. 
Combining $R_{\rm BLR}$ with the line-of-sight velocity width ($\Delta V$) measured from the variable component of the broad emission
line provides a virial black hole mass estimate as $M_{\rm BH} \equiv f {\Delta V}^2 R_{\rm BLR} /G$, 
where $G$ is the gravitational constant and $f$ is the virial factor that converts the measured virial product
into the actual black hole mass. This technique is also limited to around 50 AGNs to date since it requires extensive
photometric and spectroscopic monitoring observations. This technique has established the empirical
size-luminosity relation (Wandel et al. 1999; Kaspi et al. 2000; Bentz et al. 2006a, 2009a), which is the basis for the single-epoch (SE) method.
In the SE method, one simply substitutes the time-consuming BLR size measurement with AGN luminosity using the size-luminosity relation. This therefore provides estimates of black hole masses for broad line AGNs
from a single spectroscopic observation, thus expanding the sample size substantially at any redshift.
However, both methods suffer from the large uncertainty stemming from the unknown virial factor (see Park et al. 2012),
which depends on the geometry and kinematics of BLR of individual AGNs.

Instead, an empirically calibrated average virial factor has been applied to most AGN black hole mass
estimates, except for only a few objects where dynamical mass measurements can be obtained 
(e.g., Davies et al. 2006; Onken et al. 2007; Hicks \& Malkan 2008).
The first calibration of the virial factor was performed by Onken et al. (2004).
They derived $\langle f \rangle=5.5\pm1.8$ based on a sample of $14$ AGNs, for which both reverberation masses and 
stellar velocity dispersions were available, 
by forcing the AGN host galaxies to obey the same \msigma~relationship as for quiescent galaxies.
By enlarging the dynamical range of the AGN sample, Woo et al. (2010) determined the virial factor 
as $\log \langle f \rangle = 0.72^{+0.09}_{-0.10}$ (i.e., $\langle f \rangle = 5.2\pm1.2$) 
based on an updated reverberation sample of 24 AGNs, which included 8 low-mass local 
Seyfert 1 galaxies from the Lick AGN Monitoring project (Bentz et al. 2009b). 
They provided the upper limit of uncertainty in
the derived virial factor as 0.43 dex based on the intrinsic scatter in the relation.
In contrast, Graham et al. (2011) reported $\langle f \rangle = 2.8^{+0.7}_{-0.5}$, based on their updated
\msigma~relation of quiescent galaxies, and an upated AGN sample, which is a factor of 2 smaller than the aforementioned values. Graham et al. (2011) commented that the value of $\langle f \rangle$ might be even further lowered due to the effect of radiation pressure
(see Marconi et al. 2008).
This correspondingly reduces the black hole mass estimates for most AGNs by that amount,
influencing all of the studies incorporating single-epoch AGN black hole masses.
Thus it is important to investigate the origin of this difference and check for possible biases 
in the calibration process.

Since the derived \msigma~relation of quiescent galaxies is used to calibrate the virial factor in AGN mass estimators,
under the assumption that the same \msigma~relation holds for AGN host galaxies, 
it is important to investigate the differences in the \msigma~relations of quiescent galaxies reported in the literature, 
and to study their effect on the derived virial factors.
Originally the slopes of the \msigma~relation reported by Ferrarese \& Merritt (2000) and Gebhardt et al. (2000) were
$4.8\pm0.5$ and $3.75\pm0.3$ based on 12 and 26 galaxy samples, respectively.
After that, various slopes have been reported in the literature, ranging from $3.68$ to $5.95$.
Although the slopes are roughly consistent with the theoretical expectations of 
$M\propto\sigma^5$ (Silk \& Rees 1998) and $M\propto\sigma^4$ (Fabian 1999),
their difference and change are noteworthy.
The possible origin of the difference in slopes has been investigated in the literature.
The related factors are:
(1) the type of regression method adopted (Tremaine et al. 2002; Novak et al. 2006; see Kelly 2007 for general applications of regression),
(2) the size of the assigned uncertainty on the velocity dispersion (Merritt \& Ferrarese 2001; Tremaine et al. 2002),
(3) the velocity dispersion measures used (Tremaine et al. 2002),
(4) the adopted value of velocity dispersion for the Milky Way (Merritt \& Ferrarese 2001; Tremaine et al. 2002),
(5) the spatial resolution of the data for the resolved BH sphere-of-influence (Ferrarese \& Ford 2005; see also G\"ultekin et al. 2009, 2011; Batcheldor 2010)
(6) the morphological type of the sample used 
(Hu 2008; Graham 2008; G\"ultekin et al. 2009; Greene et al. 2010; Graham et al. 2011).

To understand the origin of the differences in the derived \msigma~relationships,
we investigate in this work 3 main issues: the difference in samples, the difference in regression methods, and the direction of the regression analysis.
Recently, with new measurements and improved modeling the number of dynamical mass measurements is continuously growing both at the high-mass and low-mass end regimes.
To date, a total of 67 black hole masses in quiescent galaxies has been measured via stellar/gas/maser kinematics
(see the most recent compilation from McConnell et al. 2011 and references therein).
Therefore, it is presently a good time to investigate what effect the difference in samples has on the derived \msigma~relation using the largest sample ever.
In addition, various estimators have been used for the regression analysis in the black hole scaling relation studies:
\texttt{FITEXY} (e.g., Tremaine et al. 2002, Novak et al. 2006, Kim et al. 2008, Li et al. 2011, Beifiori et al. 2012, Vika et al. 2012, McConnell et al. 2011),
\texttt{BCES} (e.g., Ferrarese \& Merritt 2000, Ferrarese \& Ford 2005, Hu 2008, Bentz et al. 2009a, Bennert et al. 2010, Graham et al. 2011),
\texttt{Maximum likelihood} (e.g., G\"ultekin et al. 2009, Greene et al. 2010, Schulze \& Gebhardt 2011),
Bayesian approach (\texttt{linmix\_err}) (e.g., Sani et al. 2011, Xiao et al. 2011, Mancini \& Feoli 2012).
Thus, in order to investigate the difference in the derived scaling relationships caused by the sample selection, it is important to investigate differences between the estimators for the \msigma~relation analysis.
Finally, adopting the choice of independent variable is another issue for determining the \msigma~relation.
Motivated by the suggestion by Graham et al. (2011) to use the `inverse' fit to calibrate the single-epoch AGN mass estimates, we present results based on both of the forward and inverse regressions.

This paper is organized as follows.
In the next section, we describe the most commonly used regression methods in astronomy with their explicit implementations.
In Section~\ref{sec:msigma}, we re-measure the \msigma~relation using 3 different samples from the literature
and investigate the difference due to the regression methods and samples.
In Section~\ref{sec:virial_factor} we present our main result for the calibrated virial factors and 
discuss the difference based on the regression methods and samples. The difference from the direction of regression
is discussed in Section~\ref{sec:inverse}.
Finally, we summarize and conclude in Section~\ref{sec:conclusion}.

\section{Linear regression techniques} \label{method}

Linear regression methods\footnote{For recent reviews, please see Hogg et al. (2010) and Caimmi (2011a,b).} 
in astronomy were exhaustively discussed in the pioneering paper, Isobe et al. (1990).
They provided formulae for 5 unweighted bivariate linear regression coefficients with their error estimates,
and recommended the bisector line for the case of treating the variables symmetrically.
The second paper in the series, Feigelson \& Babu (1992), extended their work by accommodating bootstrap and 
jackknife resampling procedures for error estimation, weighted regression, and truncated/censored regressions. In addition, they suggested practical strategies for linear regression problems in astronomy.

The \texttt{BCES} estimator (Bivariate Correlated Errors and intrinsic Scatter) was proposed by Akritas \& Bershady (1996) 
in order to incorporate, heteroscedastic measurement errors, intrinsic scatter, and correlation in the measurement errors. 
The method of minimizing a $\chi^2$ statistic (\texttt{FITEXY}), which account for measurement error in both the dependent and independent variable, was modified by Tremaine et al. (2002) to incorporate intrinsic scatter. They added the unknown constant intrinsic variance term in quadrature
to the error of the dependent variable and determined it so that the reduced $\chi^2$ is equal to a value of unity.
Based on the Monte Carlo simulations performed by Tremaine et al. (2002) and Novak et al. (2006), 
they concluded that the modified \texttt{FITEXY} is
a better estimator than the \texttt{BCES}. In particular, Tremaine et al. (2002) concluded that the \texttt{BCES} tends to be biased when the sample size is small or 
the mean square of the x errors is comparable to the variance of x distribution,
and that it becomes inefficient when there is a single measurement with much larger error than others.

Kelly (2007) developed a sophisticated Bayesian linear regression technique, termed \texttt{linmix\_err}. 
It accounts for intrinsic scatter in the relationship, heteroscedastic measurement errors in both the independent and
dependent variables, and correlation between the measurement errors.
This method uses a Gaussian mixture model for
the distribution of independent variables, which is shown to work well particularly when the
measurement errors are large by avoiding the bias incorporated if the choice of x-distribution model is incorrect
(also noted in Auger et al. 2010). The method assumes that the measurement errors and intrinsic scatter are Gaussian, and it accommodates multiple independent variables, nondetections, and selection effects.

Recently, G\"ultekin et al. (2009) applied a \texttt{maximum likelihood} method to determine 
the $M-\sigma$ and $M-L$ relations by naturally incorporating an intrinsic scatter and upper limits.
They also extensively investigated the distributional forms for the measurement error and intrinsic scatter.
However, they did not include a model for the distribution of the independent variable, but instead used Monte Carlo sampling to assess the impact of the measurement errors in the independent variable on the parameter estimates.

To sum up, the four methods for linear regression that have been used to characterize the black hole/host galaxy scaling relationships are:
%
%
(1) \texttt{BCES} (Akritas \& Bershady 1996),
(2) \texttt{FITEXY} (Tremaine et al. 2002),
(3) \texttt{Maximum likelihood} (G\"ultekin et al. 2009),
(4) \texttt{LINMIX\_ERR} (Kelly 2007).
In this section we explicitly show our implementation and usage of each method.
We assume the model of $y = \alpha + \beta x$ in the following analysis.

\subsection{BCES estimator}
The \texttt{BCES}($Y|X$) estimator is implemented using the formula described in Akritas \& Bershady (1996),
\begin{eqnarray}
\beta  &=& \frac{{{\mathop{\rm cov}} (x,y) - \left\langle {{\sigma _{xy}}} \right\rangle }}{{{\mathop{\rm var}} (x) - \left\langle {\sigma _x^2} \right\rangle }},\label{method:BCES:slope}\\
\alpha  &=& \left\langle y \right\rangle  - \beta \left\langle x \right\rangle ,
\end{eqnarray}
where ${\mathop{\rm cov}} (x,y)$ is the covariance of $x$ and $y$,
$\sigma_{x}$ is the standard deviation of the measurement error (i.e., standard measurement error) in $x$,
${\mathop{\rm var}} (x)$ is the 
variance of $x$, and $\sigma _{xy}$ is the covariance between the measurement errors in $x$ and $y$.
The intrinsic variance (i.e., variance in the intrinsic scatter) is estimated following the expression given in Cheng \& Riu (2006) and Kelly (2007),
\begin{equation}
{\sigma _{{\mathop{\rm int}} }} = \sqrt {{\mathop{\rm var}} (y) - \left\langle {\sigma _y^2} \right\rangle  - \beta \left[ {{\mathop{\rm cov}} (x,y) - \left\langle {{\sigma _{xy}}} \right\rangle } \right]} .
\end{equation}
The uncertainties in the parameters can be estimated with the bootstrap or using analytical estimates given in Akritas \& Bershady (1996).
In this work we assume $\sigma _{xy} = 0$ (i.e., uncorrelated measurement errors),
as most values of $x$ and $y$ in the \msigma~samples were independently measured and the covariances between the measurement errors are not provided in the literature. Thus, simply assuming the zero covariance seems to be more reasonable for these very heterogeneously collected \msigma~samples. In addition, there is no result incorporating the correlated measurement errors (i.e., $\sigma _{xy}$) to \msigma~fitting in the literature at least to our knowledge. Thus, to compare consistently with the results from the literature we here set $\sigma _{xy} = 0$.

\subsection{FITEXY estimator}
The \texttt{FITEXY} (Press et al. 1992), modified by Tremaine et al. (2002), is implemented in our work in IDL
using the \texttt{mpfit} (Markwardt 2009) Levenberg-Marquardt least-squares minimization routine.
Note that our implementation is basically similar to that given in Williams et al. (2010)
\footnote{http://purl.org/mike/mpfitexy}.
It performs the linear regression by minimizing
\begin{equation}\label{method:chi2}
{\chi ^2} = \sum\limits_{i = 1}^N {\frac{{{{\left( {{y_i} - \alpha  - \beta {x_i}} \right)}^2}}}{{\sigma _{y,i}^2 + {\beta ^2}\sigma _{x,i}^2 + \sigma _{{\mathop{\rm int}} }^2}}},	
\end{equation}
where $\alpha$ and $\beta$ are the regression coefficients,
$\sigma_{x}$ and $\sigma_{y}$ are the standard deviation in the measurement errors,
and $\sigma^2_{{\mathop{\rm int}}}$ is the intrinsic variance. The value of $\sigma_{{\mathop{\rm int}}}$ is iteratively adjusted as an effective additional $y$ error
by repeating the fit until one obtains $\chi^2/(N-2)=1$
(i.e., following the suggested iterative procedure given in Bedregal et al. 2006 and Bamford et al. 2006).
If after the initial iteration the reduced $\chi^2$ is less than one, then no further iterations
occur and one sets $\sigma_{{\mathop{\rm int}}}=0$.
We estimate uncertainties in the regression parameters with the bootstrap method.

\subsection{Maximum Likelihood estimator}\label{method:MLE}
The method of \texttt{maximum likelihood} is implemented similarly as given in 
G\"ultekin et al. (2009) (see also Woo et al. 2010).
Under the assumptions of uncorrelated Gaussian measurement errors in both coordinates and
Gaussian intrinsic scatter along $y$,
the Gaussian likelihood function is given by
\begin{equation}
{\mathcal{L}} = \prod\limits_{i = 1}^N {\frac{1}{{\sqrt {2\pi \sigma _i^2} }}} \exp \left[ { - \frac{{{{\left( {{y_i} - \alpha  - \beta {x_i}} \right)}^2}}}{{2\sigma _i^2}}} \right],
\label{eq:MLE_L}
\end{equation}
where
\begin{equation}
\sigma _i^2 = \sigma _{y,i}^2 + {\beta ^2}\sigma _{x,i}^2 + \sigma _{{\mathop{\rm int}} }^2.
\end{equation}
Note that this likelihood function implicitly assumes that the independent variable is uniformly distributed (i.e., a uniform prior for the intrinsic distribution of $x$).
Then the log-likelihood function is 
\begin{eqnarray}
- 2\ln {\mathcal{L}} &=& \sum\limits_{i = 1}^N {\ln \left( {2\pi \sigma _i^2} \right)}  + \sum\limits_{i = 1}^N {\frac{{{{\left( {{y_i} - \alpha  - \beta {x_i}} \right)}^2}}}{{\sigma _i^2}}} \nonumber \\
&=& \sum\limits_{i = 1}^N {\ln \left( {2\pi \sigma _i^2} \right)}  + {\chi ^2}.
\end{eqnarray}
This likelihood approach is more complete than the ${\chi ^2}$ method given in Equation~(\ref{method:chi2})
in a sense that it includes the intrinsic variance term in both the normalization and exponent of the likelihood
function, and determines it simultaneously with the other regression coefficients.
To estimate the best-fit parameters of $(\alpha,\beta,\sigma _{{\mathop{\rm int}} })$ using the maximum likelihood estimate (MLE), we minimize $- 2\ln {\mathcal{L}}$ using the downhill simplex method implemented as \texttt{AMOEBA}
(Press et al. 1992) in IDL.
Then we adopt uncertainties in parameters as the average difference 
where $\ln {\mathcal{L}}$ decreases from its maximum value by $0.5$.
We also estimate the parameter errors using the bootstrap method.

\subsection{Bayesian estimator (linmix\_err)} \label{method:linmix}
The Bayesian linear regression routine, \texttt{linmix\_err}, developed by Kelly (2007) is available 
from the NASA IDL astronomy user's library\footnote{http://idlastro.gsfc.nasa.gov/}.
Here we briefly summarize the method. For details, please refer to Kelly (2007).

This method assumes Gaussian intrinsic scatter, Gaussian measurement errors,
and a weighted sum of $K$ Gaussian functions for the distribution of the independent variable. The choice of a Gaussian mixture model was motivated in that it can not only approximate well various intrinsic distributions of the independent variable, but it is also a mathematically convenient conjugate family.
%
%
The full measured data likelihood function is expressed as a mixture of bivariate normal distributions
\begin{equation}
      {\mathcal{L}}
      = \prod_{i=1}^{N} \sum_{k=1}^{K} \frac{\pi_k}{2\pi |V_{k,i}|^{1/2}} 
       \exp \left[{ -\frac{1}{2} (z_i - \zeta_k)^T V_{k,i}^{-1} (z_i - \zeta_k)}\right],
       \label{eq:Bayesian_L}
\end{equation}
where $\sum_{k=1}^{K} \pi_k = 1$ and the measured data, means, and covariance matrices are, respectively,
\begin{eqnarray}
{z_i} &=& \left( {\begin{array}{*{20}{c}}
{{y_i}}\\
{{x_i}}
\end{array}} \right),\\
{\zeta _k} &=& \left( {\begin{array}{*{20}{c}}
{\alpha  + \beta {\mu _k}}\\
{{\mu _k}}
\end{array}} \right),\\
  V_{k,i} & = & \left( \begin{array}{cc}
     \beta^2 \tau_k^2 + \sigma_{{\mathop{\rm int}} }^2 + \sigma^2_{y,i} & \beta \tau_k^2 + \sigma_{xy,i} \\
     \beta \tau_k^2 + \sigma_{xy,i} & \tau_k^2 + \sigma^2_{x,i} \end{array} \right).
\end{eqnarray}
In order to calculate the posterior probability distribution of the model parameters for the given measured data,
it adopts uniform prior distributions on the regression parameters $(\alpha,\beta,\sigma _{{\mathop{\rm int}}}^2)$. It also adopt a Dirichlet, normal, and scaled inverse-$\chi^2$ prior on the mixture model parameters
$(\pi_k,\mu_k,\tau_k^2)$, respectively.
The data is `fit' using a Markov Chain Monte Carlo (MCMC) sampler.
For each regression parameter, we take the best-fit value and uncertainty as the posterior median and posterior
standard deviation from the marginal posterior distributions using the 100,000 random draws returned by the MCMC sampler.
Note that the likelihood function given in Equation~(\ref{eq:Bayesian_L}) converges to that of the 
\texttt{maximum likelihood} method given in Equation~(\ref{eq:MLE_L}) if the distribution for
the independent variable is assumed to be uniform rather than a mixture of normals and 
the measurement errors are uncorrelated.

As an illustration, we also used the likelihood function in the case of a single Gaussian model ($K=1,\pi_k=1$)
for the distribution of independent variable with uncorrelated measurement errors ($\sigma_{xy,i}=0$).
For comparison with the procedure assuming a uniform intrinsic distribution (MLE) given in Section~\ref{method:MLE}, we compute the maximum likelihood estimate (i.e., MLE$_{\rm 1G}$) utilizing 
the likelihood function derived from assuming the distribution of independent variable is a Gaussian.

In the following sections, we determine the $\alpha,\beta$, and $\sigma_{{\mathop{\rm int}}}$ parameters 
with the corresponding error estimates for the \msigma~relation using each regression technique described above.

\begin{figure} 
\centering 
\includegraphics[width=7.75cm]{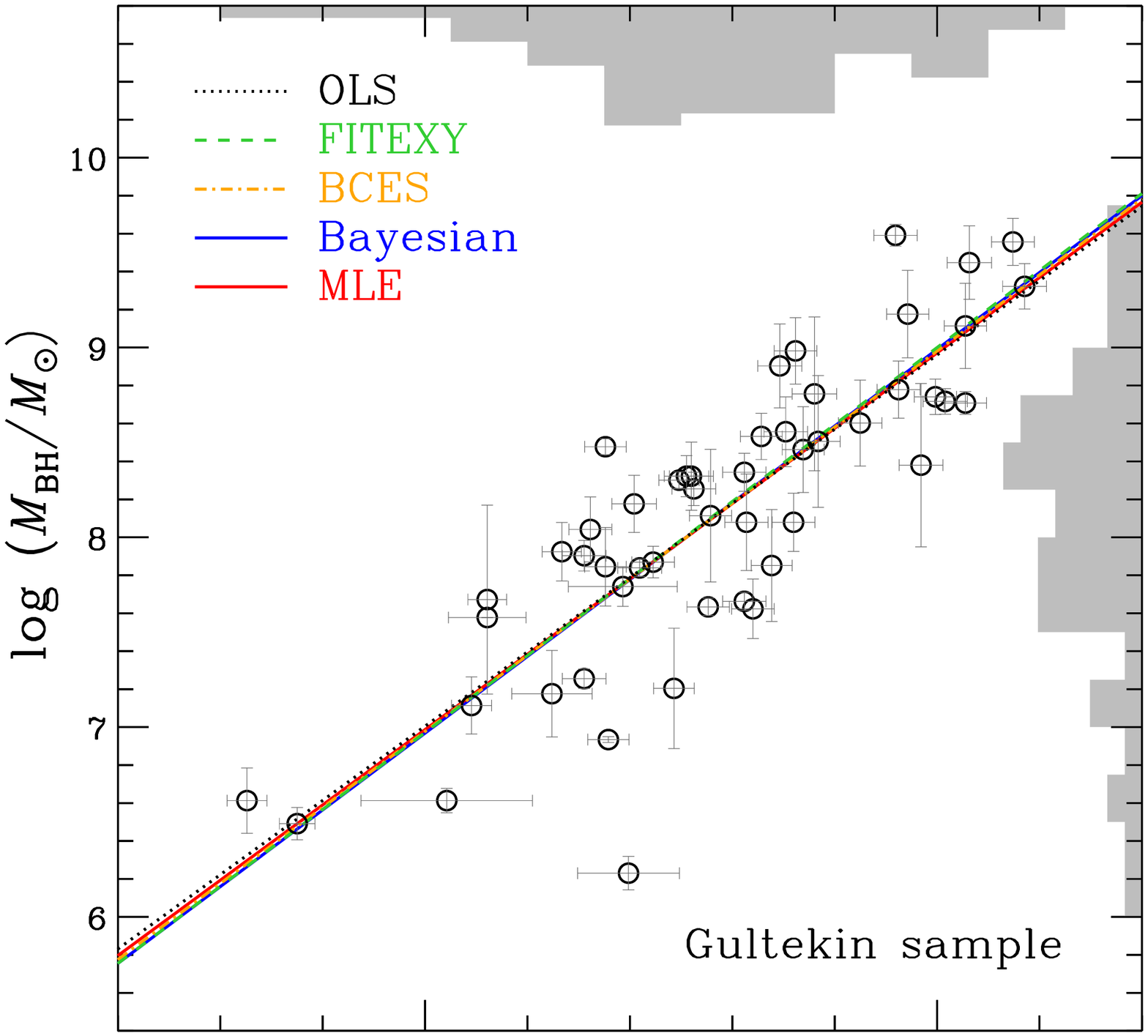}\\
\includegraphics[width=7.75cm]{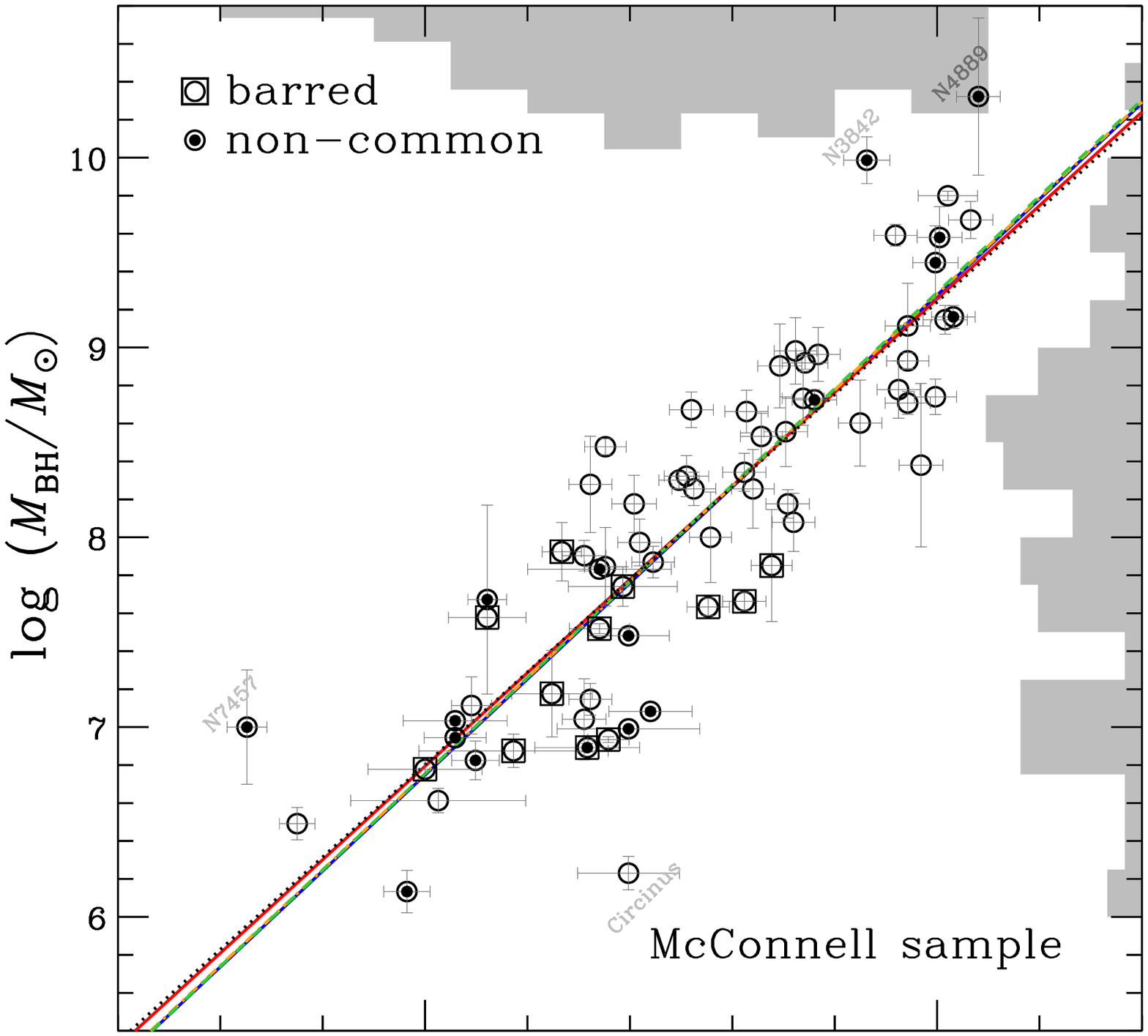}\\
\includegraphics[width=7.75cm]{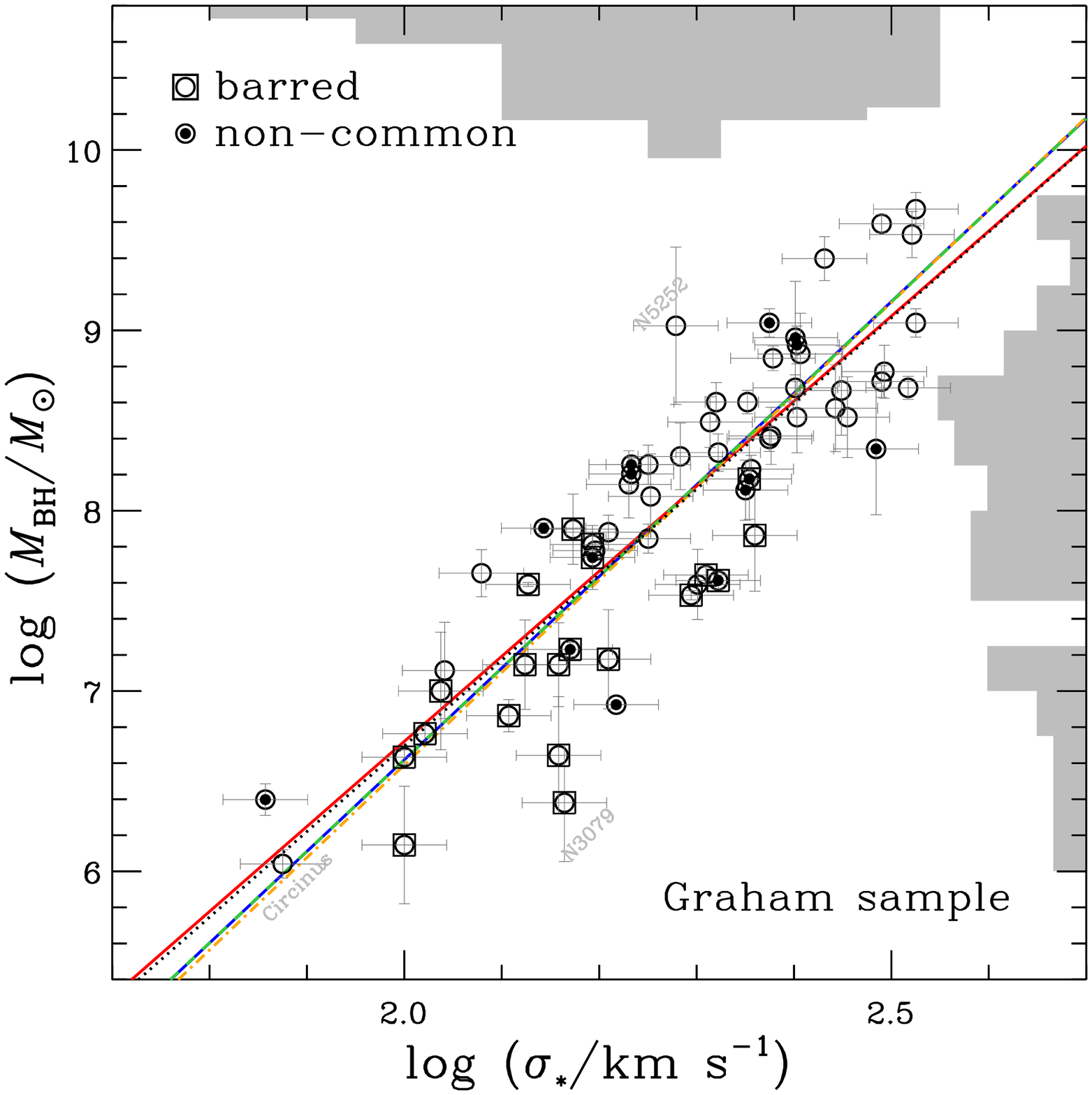}
\caption{
The \msigma~relations using each dataset from 
G\"ultekin et al. 2009 (\textit{top}),
McConnell et al. 2011 (\textit{middle}), and
Graham et al. 2011 (\textit{bottom}). Each regression line is derived from five different methods 
(see the text and Table~\ref{tab:re-estimate}).
The sample distributions for the logarithms of black hole masses and stellar velocity dispersions are 
shown in the right side and top side of each panel as grey histograms.
The non-overlapping sample in between the McConnell and Graham data are marked with a filled dot 
inside open circles. Note that the common sample has 50 galaxies.
The barred galaxies are marked with an open square enclosing open circles.
}
\label{msigma_reproduce}
\end{figure}

\section{The \msigma~relations}\label{sec:msigma}
The \msigma~relation is generally expressed as the log-linear form,
\begin{equation}
\log (M_{\rm BH} / M_{\odot}) = \alpha + \beta \log (\sigma_{\ast} / 200\ \mathrm{km~s^{-1}}) .
\label{eq:msigma}
\end{equation}
Here $y = \log (M_{\rm BH} / M_{\odot})$ and $x = \log (\sigma_{\ast} / 200\ \mathrm{km~s^{-1}})$.
%
We assume that the measurement errors in the logarithms of mass and stellar velocity dispersion
are symmetric by taking the symmetric interval in log space, i.e., 
${\epsilon _{\log {M_{{\rm{BH}}}}}} = \left( {\log M_{{\rm{BH}}}^{\rm upper} - \log M_{{\rm{BH}}}^{\rm lower}} \right)/2$ 
and ${\epsilon _{\log {\sigma _*}}} = \left( {\log \sigma_{*}^{\rm upper} - \log \sigma _{*}^{\rm lower}} \right)/2$.
%
Following Graham et al., for their data we assume that the measurement errors on the logarithm of mass are symmetric
by taking the average of upper and lower 1$\sigma$ uncertainties on the linear scale and propagating
it onto the logarithmic scale, i.e., ${\epsilon _{\log {M_{{\rm{BH}}}}}} = 0.5\left( {\epsilon _{{M_{{\rm{BH}}}}}^{\rm upper} + \epsilon _{{M_{{\rm{BH}}}}}^{\rm lower}} \right)/\left( {{M_{{\rm{BH}}}}\ln 10} \right)$. The measurement errors in the logarithm of the dispersion are 
${\epsilon _{\log {\sigma_{{*}}}}} = 0.5\left( {\epsilon _{{\sigma_{{*}}}}^{\rm upper} + \epsilon _{{\sigma_{{*}}}}^{\rm lower}} \right)/\left( {{\sigma_{{*}}}\ln 10} \right)$.
However, note that this choice of error bars does not significantly affect our results.

\subsection{Re-measuring the relation with four methods}\label{sec:re-estimate}

\begin{deluxetable*}{lcccc|cccc}
\tablecolumns{9}
\tablewidth{425pt}
\tablecaption{Re-estimation of Parameters for the \msigma~Relation of Quiescent Galaxy Samples: 
$\log (M_{\rm BH} / M_{\odot}) = \alpha + \beta \log (\sigma_{\ast} / 200\ \mathrm{km~s^{-1}})$
\label{tab:re-estimate}}
\tablehead{
\colhead{Method} & 
\colhead{} &
\colhead{Forward regression} &
\colhead{} &
\colhead{ } &
\colhead{ } &
\colhead{} &
\colhead{Inverse regression\tablenotemark{a}} &
\colhead{} \\
\colhead{} &
\colhead{} &
\colhead{} &
\colhead{} &
\colhead{} &
\colhead{} &
\colhead{} &
\colhead{} &
\colhead{} \\
\colhead{} &
\colhead{$\alpha$}    & 
\colhead{$\beta$}       & 
\colhead{$\sigma_{{\mathop{\rm int}}}$} & 
\colhead{ } &
\colhead{ } &
\colhead{$\alpha=-\alpha_{\rm inv}/\beta_{\rm inv}$}    & 
\colhead{$\beta=1/\beta_{\rm inv}$}       & 
\colhead{$\sigma_{{\mathop{\rm int}}}=\sigma_{{\mathop{\rm int, inv}}}/\beta_{\rm inv}$}
}
\startdata
& \multicolumn{8}{c}{G\"ultekin et al. (2009) Sample\tablenotemark{b}} \\
\\
OLS       & $8.18\pm0.06$ & $3.91\pm0.28$ & \nodata         & & &   $8.21\pm0.07$ & $5.60\pm0.68$ & \nodata        \\
MLE       & $8.19\pm0.06$ & $3.97\pm0.31$ & $0.39\pm0.06$   & & &   $8.22\pm0.07$ & $5.61\pm0.70$ & $0.47\pm0.08$  \\
BCES      & $8.18\pm0.06$ & $4.01\pm0.32$ & $0.38\pm0.07$   & & &   $8.20\pm0.07$ & $5.27\pm0.70$ & $0.43\pm0.10$  \\
FITEXY    & $8.19\pm0.06$ & $4.06\pm0.32$ & $0.39\pm0.06$   & & &   $8.21\pm0.07$ & $5.35\pm0.66$ & $0.45\pm0.09$  \\
Bayesian  & $8.19\pm0.07$ & $4.04\pm0.40$ & $0.42\pm0.05$   & & &   $8.21\pm0.08$ & $5.44\pm0.56$ & $0.49\pm0.09$  \\
\hline
& \multicolumn{8}{c}{McConnell et al. (2011) Sample\tablenotemark{c}} \\
\\
OLS       & $8.27\pm0.06$ & $4.87\pm0.37$ & \nodata         & & &   $8.33\pm0.06$ & $6.55\pm0.50$ & \nodata        \\
MLE       & $8.28\pm0.06$ & $4.92\pm0.34$ & $0.41\pm0.05$   & & &   $8.33\pm0.06$ & $6.43\pm0.51$ & $0.47\pm0.06$  \\
BCES      & $8.28\pm0.06$ & $5.06\pm0.41$ & $0.43\pm0.05$   & & &   $8.33\pm0.06$ & $6.36\pm0.51$ & $0.48\pm0.07$  \\
FITEXY    & $8.28\pm0.06$ & $5.07\pm0.36$ & $0.43\pm0.05$   & & &   $8.32\pm0.06$ & $6.29\pm0.49$ & $0.47\pm0.06$  \\
Bayesian  & $8.27\pm0.06$ & $5.06\pm0.36$ & $0.44\pm0.05$   & & &   $8.32\pm0.07$ & $6.31\pm0.46$ & $0.49\pm0.07$  \\
\hline
& \multicolumn{8}{c}{Graham et al. (2011) Sample} \\
\\
OLS       & $8.13\pm0.05$ & $4.75\pm0.29$ & \nodata         & & &   $8.16\pm0.06$ & $6.22\pm0.46$ & \nodata        \\
MLE       & $8.14\pm0.05$ & $4.72\pm0.29$ & $0.30\pm0.04$   & & &   $8.17\pm0.06$ & $6.06\pm0.46$ & $0.33\pm0.05$  \\
BCES      & $8.13\pm0.05$ & $5.13\pm0.35$ & $0.31\pm0.04$   & & &   $8.15\pm0.06$ & $5.95\pm0.45$ & $0.34\pm0.05$  \\
FITEXY    & $8.15\pm0.05$ & $5.08\pm0.34$ & $0.31\pm0.04$   & & &   $8.16\pm0.05$ & $5.84\pm0.42$ & $0.33\pm0.05$  \\
Bayesian  & $8.15\pm0.05$ & $5.08\pm0.36$ & $0.31\pm0.05$   & & &   $8.17\pm0.06$ & $5.85\pm0.42$ & $0.34\pm0.06$ 
\enddata
\tablecomments{
Forward regression=fit $\log M_{\rm BH}$ on $\log \sigma_*$; 
Inverse regression=fit $\log \sigma_*$ on $\log M_{\rm BH}$; 
OLS=Ordinary Least Squares; MLE=Maximum Likelihood Estimates;
BCES=estimator of Akritas \& Bershady (1996); FITEXY=estimator of Tremaine et al. (2002); 
Bayesian=Bayesian posterior median estimates using \texttt{linmix\_err} procedure of Kelly (2007).
}
\tablenotetext{a}{Inverse regression and its results will be discussed in Section~\ref{sec:inverse}.}
\tablenotetext{b}{We used 49 galaxies listed in Table 1 in G\"ultekin et al. (2009) 
without upper limits for comparison with other samples. 
For the MLE estimate, we also estimate the parameters using the same likelihood
function and error estimation method given in G\"ultekin et al. (2009).
The result is $(\alpha, \beta, \sigma_{\mathop{\rm int}}) = (8.18\pm0.06, 3.97\pm0.39, 0.42\pm0.05)$.
Note that there are two different mass measurements for NGC~1399 and NGC~5128.}
\tablenotetext{c}{We used all of the 67 galaxies listed in Table 4 in McConnell et al. (2011), 
while they used only 65. We found that there is a typo in the $M_{\rm BH}$ of NGC~1023 in their Table 4.
Thus we corrected the value from $14.6\times10^7$ to $4.6\times10^7$.
Note that there are two different mass measurements for the NGC~1399 and for the NGC~5128.
Following their scheme, if we apply half weights for them, then we get the same result with that of
McConnell et al. (2011), i.e., $(\alpha, \beta, \sigma_{\mathop{\rm int}}) = (8.28\pm0.06, 5.13\pm0.34,
0.42\pm0.05)$ for the FITEXY estimate.}
\end{deluxetable*}
In this section we consistently re-derive the \msigma~relation using three literature samples to check our 
implementation of the fitting methods.
Figure~\ref{msigma_reproduce} shows the re-estimated \msigma~relations of the data from G\"ultekin et al. 2009
(\textit{top}), McConnell et al. 2011 (\textit{middle}), and Graham et al. 2011 (\textit{bottom})
using the various methods described in the previous section. We also include the ordinary least squares (hereafter  \texttt{OLS}) line as a reference.
%
For the sample of 49 galaxies without upper limits from Table 1 in G\"ultekin et al. (2009),
we also follow the same fitting scheme of their maximum likelihood estimator, 
which is slightly different to the method described in Section~\ref{method:MLE}. Using the G\"ultekin et al. (2009) procedure, we first perform the fit without accounting for the measurement errors in the independent variables. 
Then, we incorporate the effects of measurement errors in $x$ into the parameter uncertainties by adding in quadrature the standard deviations estimated from the Monte Carlo fitting results for $10^{4}$ realizations of the independent variables.
%
Recently, McConnell et al. (2011) have updated the compiled sample of 49 galaxies from G\"ultekin et al. (2009)
by including new black hole mass measurements and revising earlier black hole masses based on improved stellar
orbit modeling, which accounts for dark matter halos (Gebhardt \& Thomas 2009; van den Bosch \& de Zeeuw 2010; Shen \& Gebhardt 2010; Schulze \& Gebhardt 2011).
We have consistently re-estimated slopes of the \msigma~relation using the 67 galaxies listed in Table 4 of McConnell et al. (2011).
%
The independently compiled sample of 64 galaxies from Graham et al. (2011) is also used.
Table~\ref{tab:re-estimate} lists all regression results.
Note that we get consistent results with each paper if we choose the same method and setting used by the respective papers.

For the G\"ultekin et al. sample the difference between fitted lines is only marginal since they assumed a minimum of 5\% 
measurement errors on $\sigma_*$; such small errors in $\sigma_*$ are found to have relatively small impact on the regression result, as described in the next section.
However, the slope of the \msigma~relation derived from the updated sample of McConnell et al. is significantly
larger than that of the G\"ultekin et al., increasing from $\sim4$ to $\sim5$.
As implied by the histograms for the black hole mass distributions shown in grey in the figures, 
this increase is mostly due to inclusion of both the low-mass and high-mass sample,
which generally show an offset trend to the relation of the G\"ultekin et al. sample.
For the Graham et al. sample the difference between fitted lines is marginally significant,
and the slope is apparently divided into two groups, since they assigned relatively
large (10\%) measurement errors in $\sigma_*$. It seems that the bias of the \texttt{maximum likelihood} estimator
starts to become non-negligible for $x$-errors of this magnitude. Therefore we investigate in details the effect of the amplitude of the $x$-error
on these estimators in the following sections.

\subsection{The effect of the adopted measurement uncertainty of $\sigma_*$}

\begin{figure} 
\centering
\includegraphics[width=8cm]{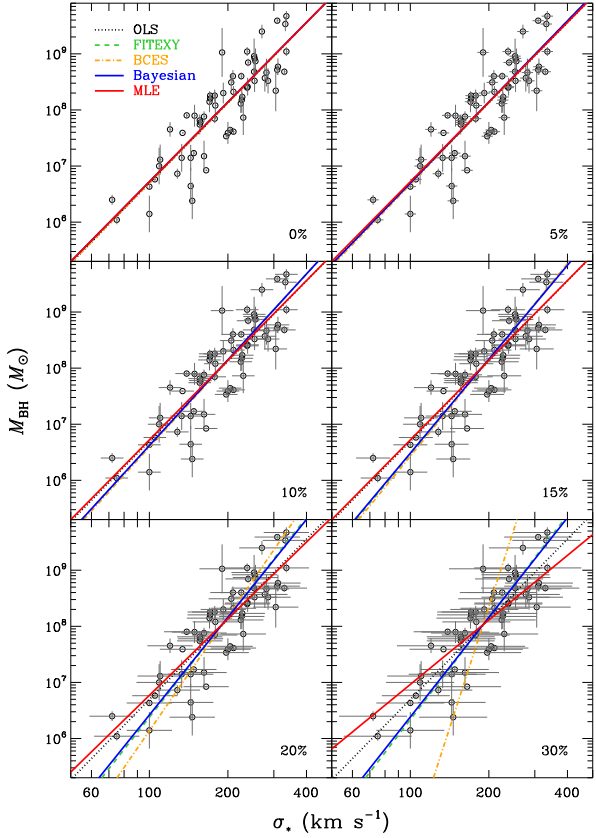} 
\caption{
Comparison of five regression lines as a function of the assigned amount of measurement errors on $\sigma_*$
using Graham et al. (2011) sample. The percentages of assigned errors are given at lower right corners in each panel.
In the case of measurement errors on independent variables above 10\%, the difference between the 
fitted lines is clearly visible.
}
\label{fig-graph}
\end{figure}

\begin{figure} 
\centering
\includegraphics[width=7cm]{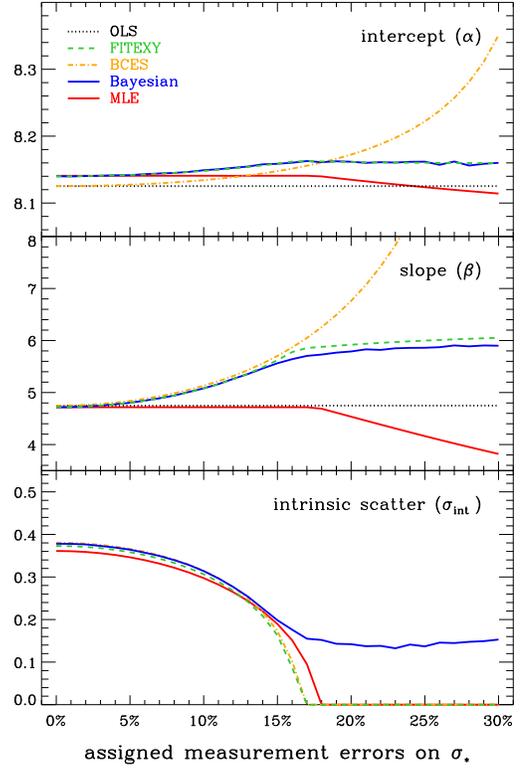} 
\caption{Direct comparison of regression results for the intercept (\textit{top}), slope (\textit{middle}),
and intrinsic scatter (\textit{bottom}) from the analysis of Figure~\ref{fig-graph}.
For the 5\% errors the difference is only marginal.
When the measurement errors are larger than 15\% there is significant deviation between the estimators.
}
\label{fig-result}
\end{figure}

Merritt \& Ferrarese (2001) first noted that ignoring the measurement errors in the velocity dispersion 
leads to a biased slope (i.e., underestimates) in the \msigma~relation. 
However, Tremaine et al. (2002) argued that the effect of the measurement errors in the velocity dispersion
is not significant even at the 10\% error level using two estimators, \texttt{BCES} and \texttt{FITEXY}, 
based on their simulation results.
Indeed, 
the measurement errors in the independent variables can have significant impact on the regression analysis 
as also investigated by Kelly (2007), although the effect is only marginal in current datasets of the \msigma~relation.
Typically the measurement errors in the velocity dispersion are assumed to be 5\% or 10\% in literature.

In Figure~\ref{fig-graph} we compare the fitted lines to the Graham et al. \msigma~dataset assuming measurement errors on $\sigma_*$ ranging from 0\% to 30\%.
The difference between the fitted lines is noticeable and obvious when the measurement errors are large.
Figure 3 compares the regression coefficients and intrinsic
scatter derived from the five estimators as a function of the assigned errors.
As a reference we show the results from the \texttt{OLS} estimator, i.e., for the unweighted
fitting scheme without accounting for the intrinsic scatter as described in Isobe et al. (1990).
This estimator is biased when there are measurement errors.

For the intercept, the estimators do not give significantly different results except for the case of the \texttt{BCES} estimator.
Both the intercept and slope estimated from the \texttt{BCES} estimator show very different behaviour 
in the high measurement error regime, which is consistent with the result from Tremaine et al. (2002). 

For the slope, it is noted that all converge to the same value as the measurement errors
in the velocity dispersion approach zero. Moreover, the estimators \texttt{BCES}, \texttt{FITEXY}, and \texttt{Bayesian} are very similar up to the 15\% error level, thus indicating consistent estimation for these three estimators.
%
The value of the slope from the \texttt{BCES} estimator becomes higher compared to the others 
as the assumed errors on $\sigma_*$ increase.
This is expected from the denominator term in Equation~(\ref{method:BCES:slope}).
As can be seen, the estimated slope from the \texttt{maximum likelihood} estimator
is almost identical to 
the \texttt{OLS} result up to measurement error amplitudes of $\sim 16\%$ on the independent variable.
As noted and discussed in Kelly (2007) this biased behaviour is due to the implicit adoption of a naive uniform prior for
the intrinsic distribution of the independent variables; this bias is also noted in K\"ording et al (2006).
Based on this, we do not recommend the \texttt{maximum likelihood} method as outlined in Section~\ref{method:MLE}.
It is surprising that the \texttt{FITEXY} with an ad hoc iterative approach gives fairly consistent results with 
that of the fully Bayesian approach (\texttt{linmix\_err}).
This is inconsistent with the result of the simulations performed by Kelly (2007). The source of this discrepancy is that for the \texttt{FITEXY} estimator Kelly (2007) did not refit the slope and intercept each time after the intrinsic scatter term is
iteratively adjusted (see also, Kelly 2011).
Instead, he just assigned the intrinsic scatter value such that the reduced $\chi^2$ is equal to unity
using the first minimization result of $\alpha$ and $\beta$ for the zero intrinsic scatter case 
(i.e., just simply increasing $\sigma_{{\mathop{\rm int}}}$ without re-optimization each time).

For the intrinsic scatter, its level is very sensitive to the magnitude of the assumed measurement errors.
Only \texttt{linmix\_err} recovers a non-zero intrinsic scatter amplitude even 
in the case of assuming large measurement errors on $\sigma_*$.

\subsection{Monte Carlo simulations}
\begin{figure*} 
\centering
\includegraphics[width=7.5cm]{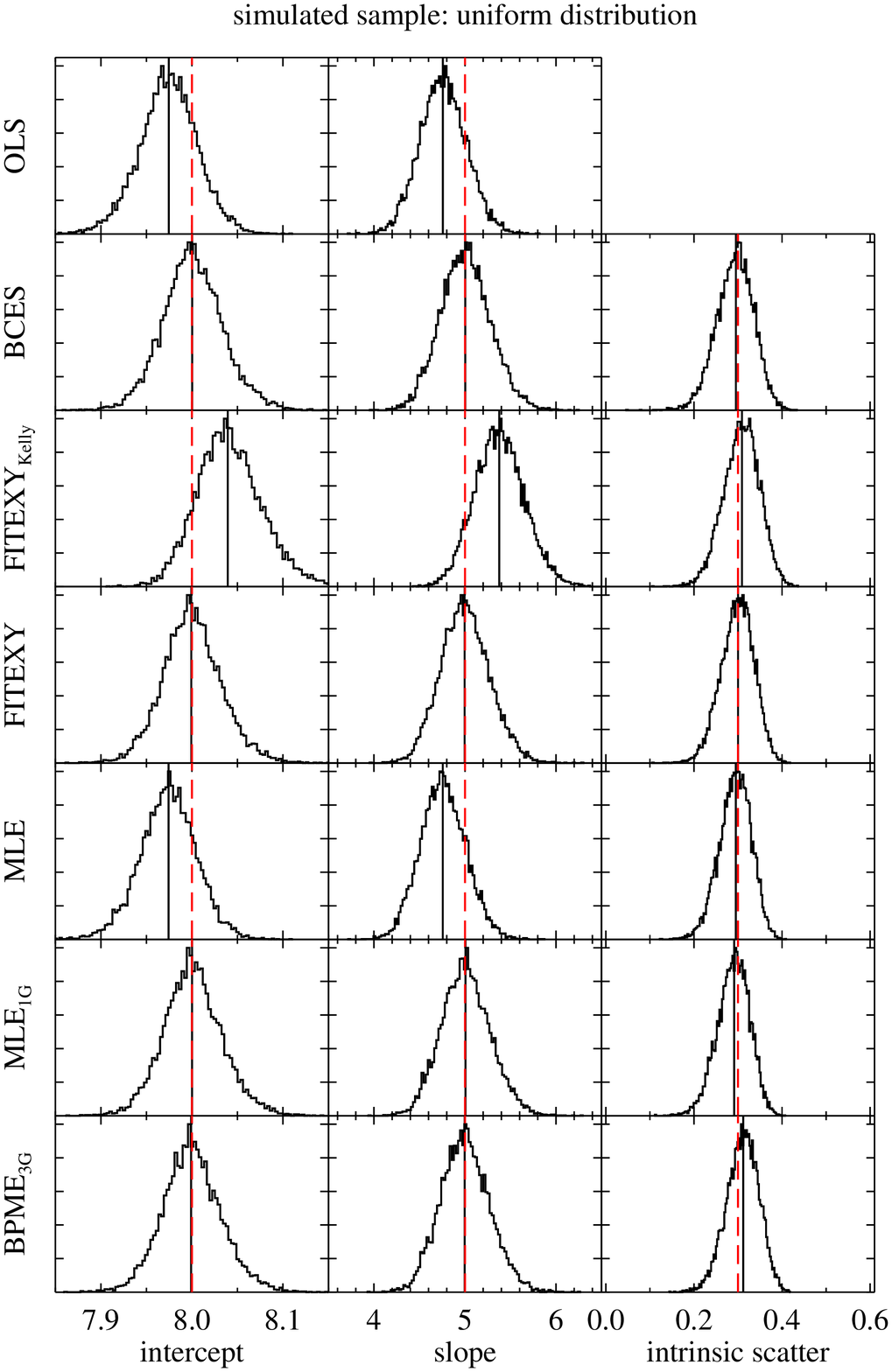} \hspace{2em}
\includegraphics[width=7.5cm]{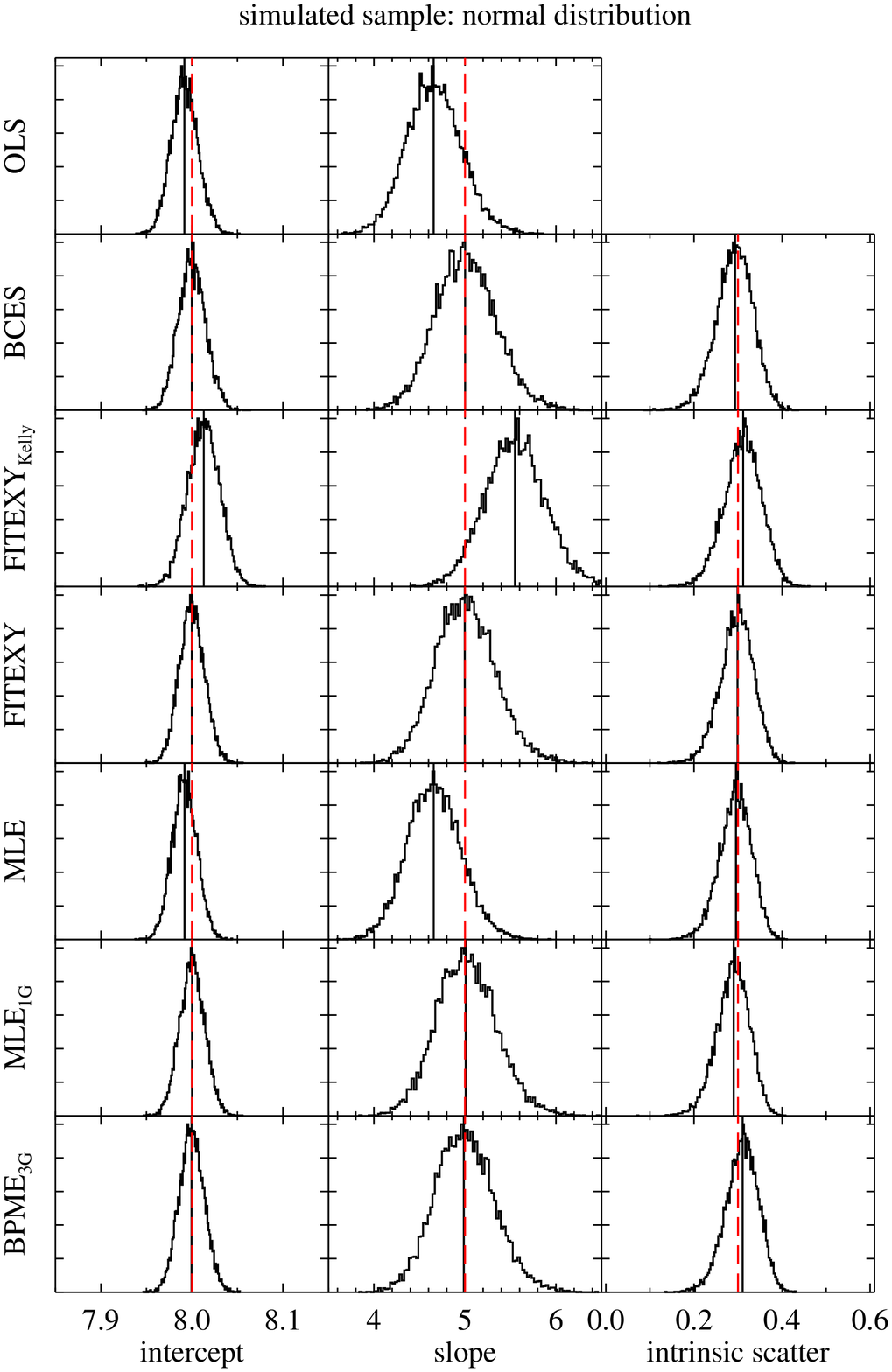} \\ \vspace{1em}
\includegraphics[width=7.5cm]{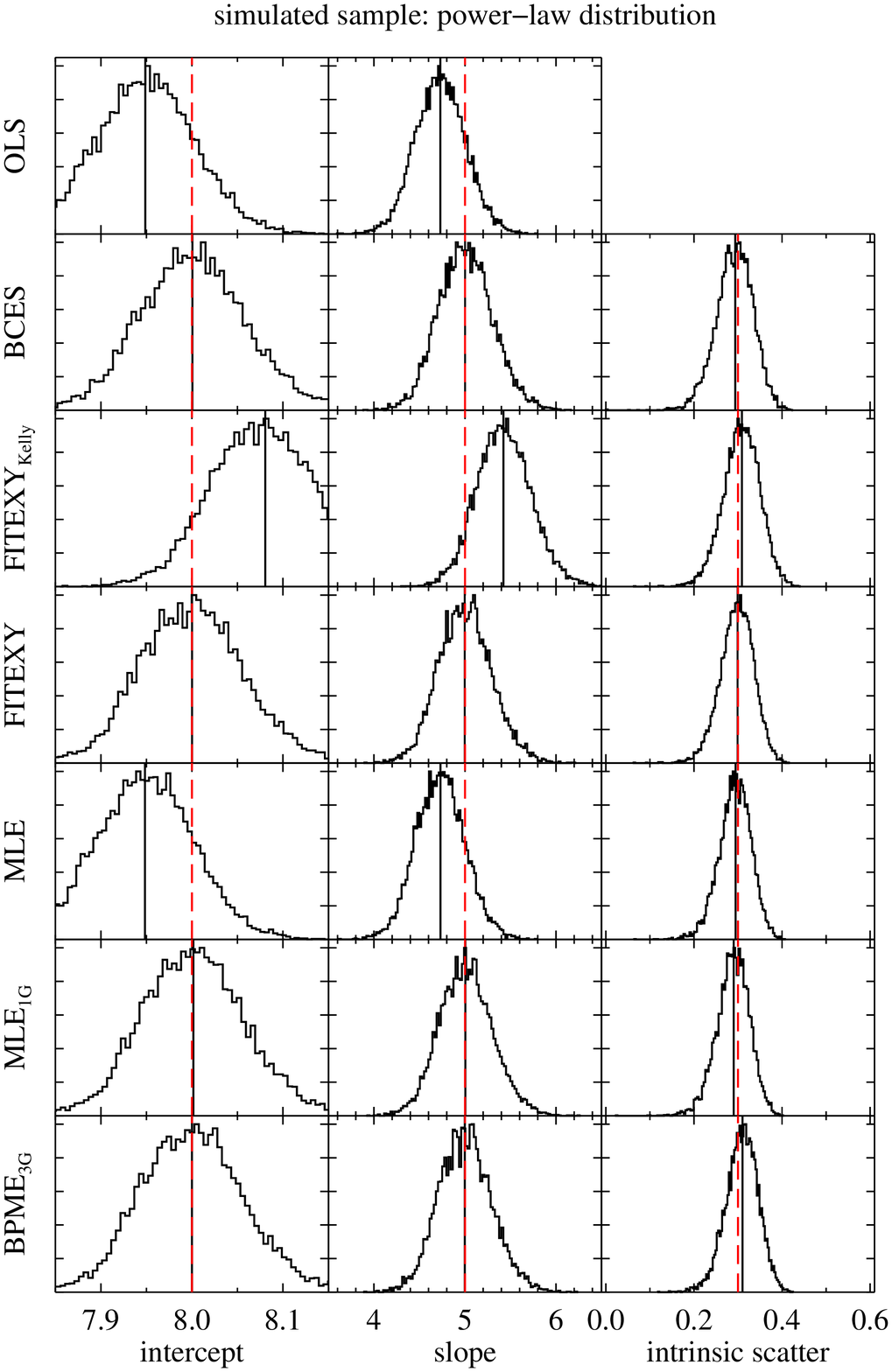} 
\caption{Monte Carlo simulation results for the cases of the given uniform (\textit{upper left}), normal (\textit{upper right}), and power-law (\textit{bottom}) distributions on $x$.
Each column shows the distribution of intercept, slope, and intrinsic scatter estimated from the simulated datasets
using the various estimators.
FITEXY$_{\rm Kelly}$ is the version of \texttt{FITEXY} estimator implemented by Kelly (2007).
MLE$_{\rm 1G}$ means the \texttt{maximum likelihood} estimator with a single Gaussian model 
for the distribution of independent variable as described in Section~\ref{method:linmix}.
BPME$_{\rm 3G}$ is the Bayesian posterior median estimate using \texttt{linmix\_err} procedure
based on the normal mixture model with 3 Gaussians.
The median value of the simulated distribution is plotted as a vertical solid black line 
while the true value is indicated by the red dashed line in each panel.}
\label{fig:MCsimul10:uniform}
\end{figure*}

An incorrect model for the distribution of the true values of $x$ and $y$ can lead to
biased slope estimates, especially in the case of relatively large measurement errors on the independent variables,
as pointed out by Gull (1989) and Kelly (2007) (see also, Auger et al. 2010, Mantz et al. 2010, and March et al. 2011).
Here we use Monte Carlo simulations to investigate the effect of an incorrect assumption on the intrinsic distribution of
the independent variables.
First, we generate three 10,000 simulated datasets by assuming respectively uniform, normal, and power-law distributions
for $x$.
The number of data points in each realization is set to be the same as that of the Graham \msigma~data set (i.e., 64).
The true intercept, slope, and standard deviation of Gaussian intrinsic scatter are assumed to be 8, 5, and 0.3 dex respectively, similar to typical values from the regression results given in Table~\ref{tab:re-estimate}. 
In other words, the sample points ($y$) from the given intrinsic relation ($y=8+5x$) are scattered by the Gaussian random offsets with $\sigma=0.3$.
Then Gaussian random noises, having zero mean and standard deviations equal to the measurement errors from the Graham et al. (2011) data set, are added to both $x$ and $y$.
We fit the simulated data sets using the regression methods described in Section~\ref{method}.

Figure \ref{fig:MCsimul10:uniform} shows 
the simulation results for the uniform, normal, and power-law distributions of $x$, respectively.
As already pointed above, the estimated intercept and slope from \texttt{MLE} are biased and distributed similarly to that of the \texttt{OLS} estimator. This bias is regardless of the form for the intrinsic distribution, and surprisingly the \texttt{MLE} still has a bias for the simulated sample from the uniform distribution. This is because the the \texttt{maximum-likelihood} method assumes a uniform distribution for the independent
variable in the range of $-\infty$ to $\infty$, while in the simulations performed here the uniformly distributed
data have some finite range (i.e., fixed to be same as the range of Graham et al. data).
As can be seen, the true values are well recovered if the likelihood function is changed to assuming a Gaussian
for the distribution of independent variables, as described in Section~\ref{method:linmix} (MLE$_{\rm 1G}$).
This modified maximum likelihood estimates give very similar distributions to that of the fully Bayesian estimates
based on the \texttt{linmix\_err} procedure (BPME$_{\rm 3G}$).
Here we also show the result of the version of \texttt{FITEXY} estimator used by Kelly (2007) to compare directly
(FITEXY$_{\rm Kelly}$).
The biased behaviour is same as noted in Kelly (2007). Thus this means that his implementation of \texttt{FITEXY}
is inefficient compared to the one implemented here.
From a viewpoint of how well the true values are recovered,
all of \texttt{BCES}, \texttt{FITEXY}, and \texttt{Bayesian} estimators performed very well in this test.
%

\begin{figure} 
\centering 
\includegraphics[width=8cm]{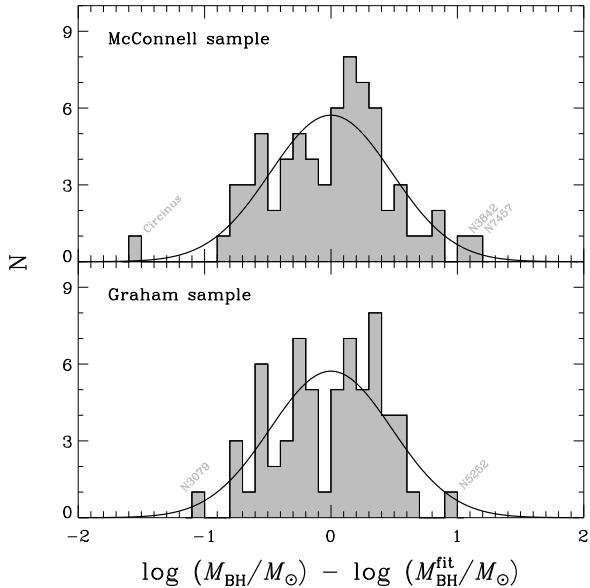}
\caption{
Histograms of mass residuals from the \msigma~relation based on the \texttt{FITEXY} estimator.
A single Gaussian fit with the center fixed to zero is expressed as a solid line.
}
\label{msigma_residual}

\end{figure}

\begin{figure} 
\centering 
\includegraphics[width=8cm]{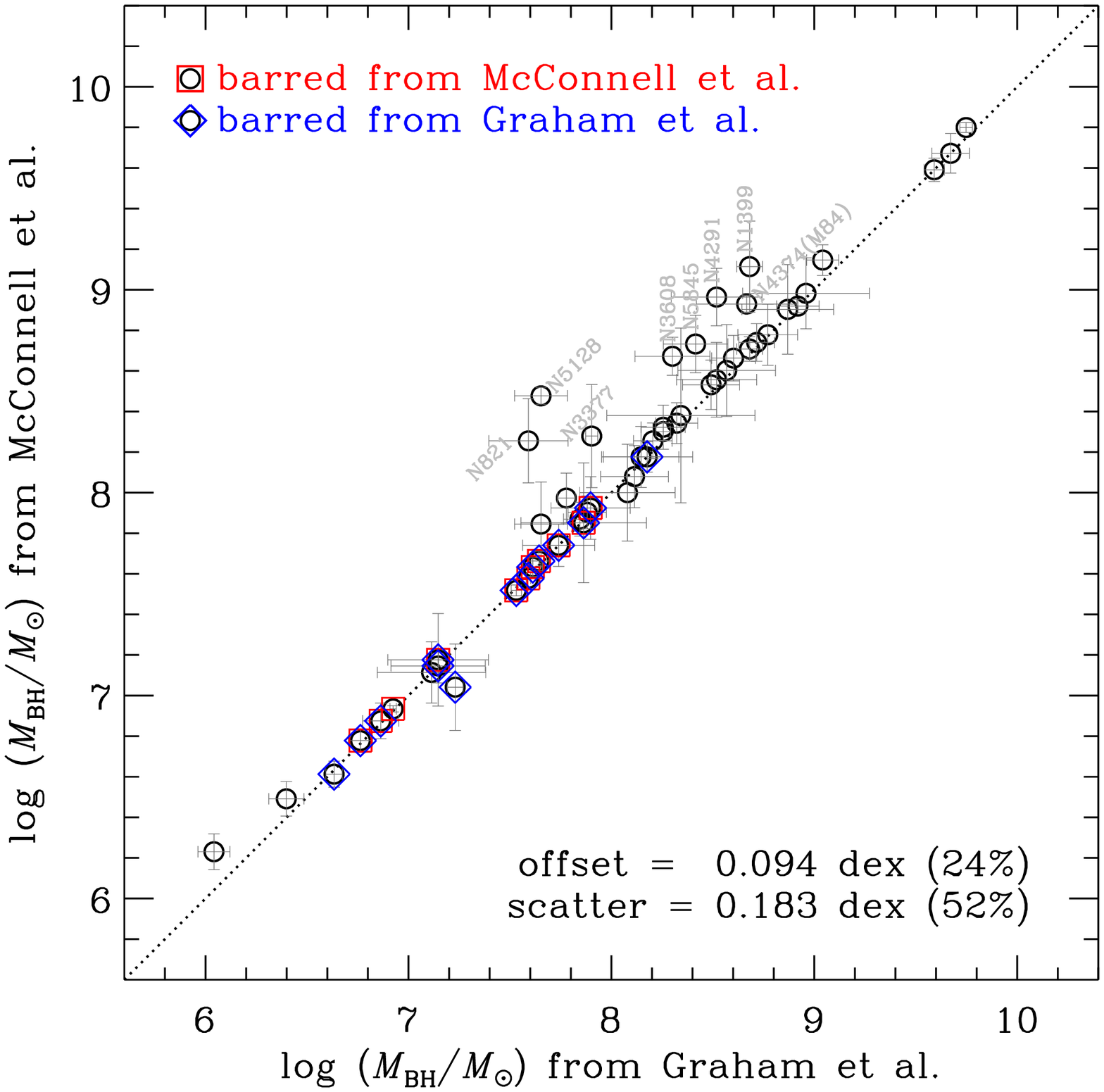}\\
\includegraphics[width=8cm]{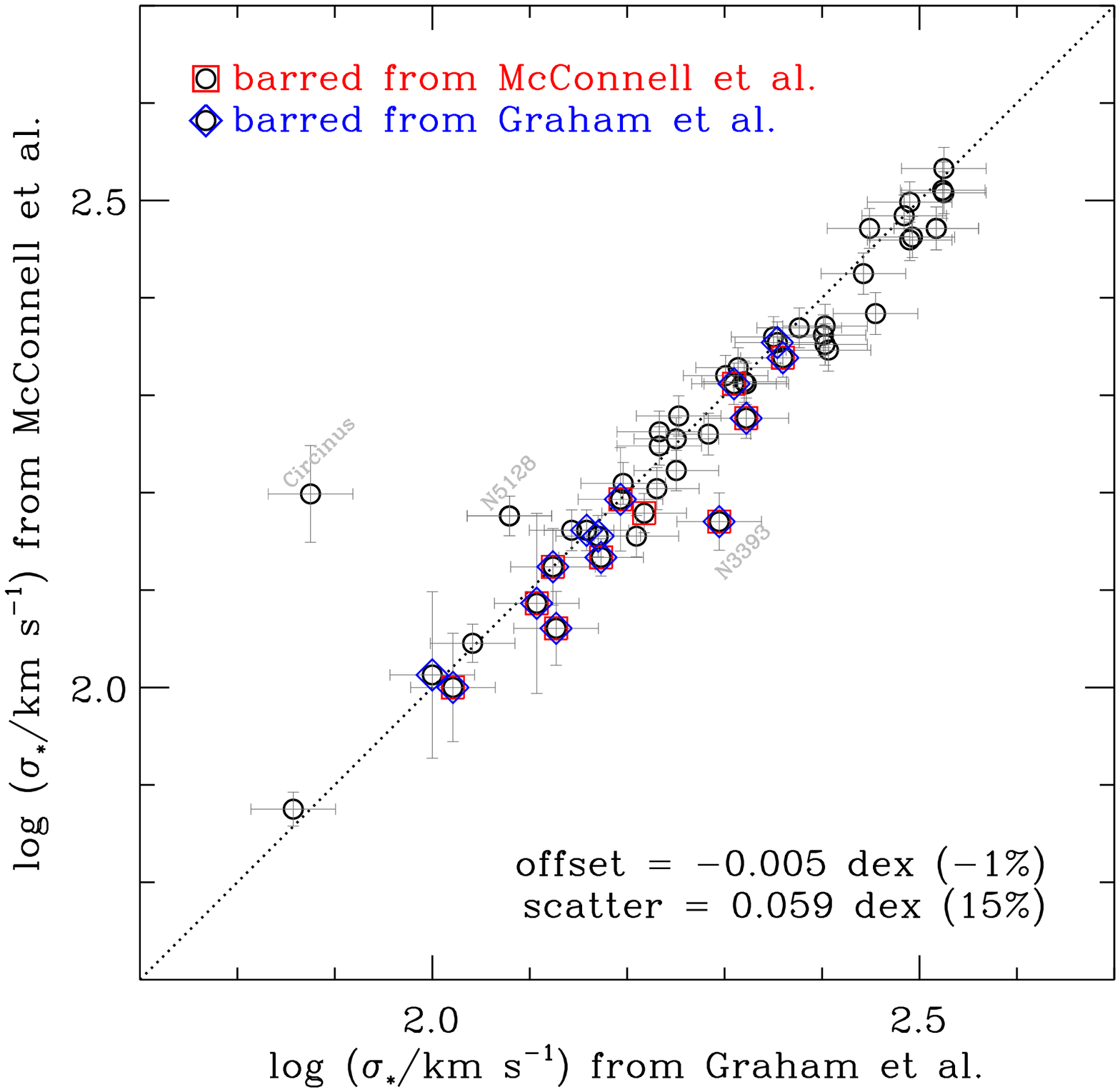}
\caption{
Comparison of the values of \mbh~and $\sigma_{*}$ for the galaxies in the set defined by the intersection of the 
McConnell et al. (2011) and Graham et al. (2011) samples.
The barred sample defined by McConnell et al. (2011) is marked with a red open square, 
while that defined by Graham et al. (2011) is marked with a blue open diamond.
The dotted line indicates the identity relationship.
}
\label{fig:compare_m_sigma_values}
\end{figure}

\subsection{The sample difference}\label{sec:sample_issues}

In this section we investigate the sample discrepancy in detail using the most updated samples of McConnell
et al. and Graham et al. We do this because the \msigma~relations derived from these two datasets show a difference in 
the intercept (see Table~\ref{tab:re-estimate}), which consequently affects the value of the virial
factor. Note that the change in the \msigma~relation from the G\"ultekin sample to the McConnell sample is obvious because there
was a major update of the sample as discussed in Section~\ref{sec:re-estimate}.
However, the difference between the McConnell and Graham data is not clear 
since these two sample have 50 galaxies in common.

Figure~\ref{msigma_residual} shows the mass residuals from the best-fit \msigma~relation derived from the McConnell and Graham data. As can be seen, there is one extreme outlier, Circinus, in the McConnell sample.
Note that the central velocity dispersions of Circinus used in McConnell (i.e., 158 km/s) and Graham (i.e., 75 km/s)
data are different from each other even though they were both taken from the HyperLEDA\footnote{http://leda.univ-lyon1.fr/} database.
The central velocity dispersion for this object currently given in the HyperLEDA is $157.6\pm18.8$ km/s.
We verified that the value listed by Graham et al. was a typo (Alister W. Graham 2012, private communication).

In order to investigate the difference in the sample in common between Graham and McConnell, in Figure~\ref{fig:compare_m_sigma_values} we compare the values of the black hole masses and stellar velocity dispersions.
For the $M_{\rm BH}$ values, there are quite a few galaxies deviating from the identity relation. 
This is mostly due to recent updates of the black hole masses by Schulze \& Gebhardt (2011) in the McConnell data.
For the $\sigma_*$, the values of the Graham sample are slightly larger on average than those of the McConnell sample, 
except for a few outliers. This slight average difference stems from the difference in the adopted velocity
dispersion measures. Graham et al. (2011) used the central velocity dispersion provided in HyperLEDA, 
while McConnell et al. (2011) mostly used the effective velocity dispersion whenever it was available.
This leads to systematic differences in the dispersion values as discussed in Tremaine et al. (2002).
These mass and dispersion differences work to make the intercept smaller in the Graham sample compared to the McConnell sample.
However, we note that the barred sample does not show any significant difference
between these datasets. 
We also performed the regression for the common sample only, and found that the intercepts remained almost the same
while the slopes were reduced by $0.2-0.3$ compared to that of the entire sample.
Therefore, the difference of the intercepts between McConnell and Graham samples is due to 
the different values adopted for the common sample, while the difference in slopes is primarily due to the non-overlapping sample.

\section{The virial factor}\label{sec:virial_factor}

\begin{deluxetable}{lcccc}
\tablecolumns{5}
\tablewidth{250pt}
\tablecaption{AGN black hole masses and stellar velocity dispersions} 
\tablehead{ 
\colhead{Galaxy} &
\colhead{VP$_{\rm BH}$} & 
\colhead{VP$_{\rm BH}$ ref.} & 
\colhead{$\sigma_{*}$} & 
\colhead{$\sigma_*$ ref.} \\ 
\colhead{} &
\colhead{($c\tau_{\rm cent}\sigma_{\rm line}^2/G$)} &
\colhead{} &
\colhead{} &
\colhead{} \\
\colhead{} &
\colhead{$10^{6}$ \msun} &
\colhead{} &
\colhead{\kms} &
\colhead{} \\
\colhead{(1)} &
\colhead{(2)} &
\colhead{(3)} &
\colhead{(4)} &
\colhead{(5)}
} 
\startdata
3C 120             & $10.1^{+5.7}_{-4.1}$   & 1       & $162\pm20$ & 2\\
3C 390.3           & $52.2^{+11.7}_{-11.7}$ & 1       & $273\pm16$ & 1\\
Ark 120            & $27.2^{+3.5}_{-3.5}$   & 1       & $221\pm17$ & 1\\
Arp 151            & $1.31^{+0.18}_{-0.23}$ & 4 \& 6  & $118\pm4$  & 6\\
Mrk 50             & $6.2^{+0.9}_{-0.9}$    & 7       & $109\pm14$ & 7\\
Mrk 79             & $9.52^{+2.61}_{-2.61}$ & 1       & $130\pm12$ & 1\\
Mrk 110            & $4.57^{+1.1}_{-1.1}$   & 1       & $ 91\pm 7$ & 3\\
Mrk 202            & $0.55^{+0.32}_{-0.22}$ & 4 \& 6  & $78\pm3$   & 6\\
Mrk 279            & $6.35^{+1.67}_{-1.67}$ & 1       & $197\pm12$ & 1\\
Mrk 590            & $8.64^{+1.34}_{-1.34}$ & 1       & $189\pm 6$ & 1\\
Mrk 817            & $11.3^{+2.7}_{-2.8}$   & 5       & $120\pm15$ & 1\\
Mrk 1310           & $0.61^{+0.20}_{-0.20}$ & 4 \& 6  & $84\pm5$   & 6\\
NGC 3227           & $1.39^{+0.29}_{-0.31}$ & 5       & $136\pm 4$ & 1\\
NGC 3516           & $5.76^{+0.51}_{-0.76}$ & 5       & $181\pm 5$ & 1\\
NGC 3783           & $5.42^{+0.99}_{-0.99}$ & 1       & $ 95\pm10$ & 4\\
NGC 4051           & $0.31^{+0.10}_{-0.09}$ & 5       & $ 89\pm 3$ & 1\\
NGC 4151           & $8.31^{+1.04}_{-0.85}$ & 2       & $ 97\pm 3$ & 1\\
NGC 4253 (Mrk 766) & $0.35^{+0.15}_{-0.14}$ & 4 \& 6  & $93\pm32$  & 6\\
NGC 4593           & $1.78^{+0.38}_{-0.38}$ & 3       & $135\pm 6$ & 1\\
NGC 4748           & $0.68^{+0.24}_{-0.30}$ & 4 \& 6  & $105\pm13$ & 6\\
NGC 5548           &$12.41^{+3.06}_{-4.21}$ & 4 \& 6  & $195\pm13$ & 6\\
NGC 6814           & $3.73^{+1.10}_{-1.11}$ & 4 \& 6  & $95\pm3$   & 6\\
NGC 7469           & $2.21^{+0.25}_{-0.25}$ & 1       & $131\pm 5$ & 1\\
PG 1426+015        & $236^{+70}_{-70}$      & 1       & $217\pm15$ & 5\\
SBS 1116+583A      & $1.08^{+0.52}_{-0.49}$ & 4 \& 6  & $92\pm4$   & 6 
\enddata
\label{tab:AGNdata}
\tablecomments{
Col. (1) name.
Col. (2) virial product (VP$_{\rm BH}=M_{\rm BH}/f$) based on 
the line dispersion ($\sigma_{\rm line}$) from reverberation mapping.
Col. (3) reference for virial product. 1. Peterson et al.\ 2004;
2. Bentz et al.\ 2006b; 3. Denney et al.\ 2006;
4. Bentz et al.\ 2009b; 5. Denney et al. 2010; 6. Park et al. 2012; 7. Barth et al. 2011.
Col. (4) stellar velocity dispersion.
Col. (5) reference for stellar velocity dispersion. 1. Nelson et al. 2004;
2. Nelson \& Whittle 1995; 3. Ferrarese et al. 2001; 4. Onken et
al. 2004; 5. Watson et al. 2008; 6. Woo et al. 2010; 7. Barth et al. 2011.
}
\end{deluxetable}

\begin{figure} 
\centering 
\includegraphics[width=8cm]{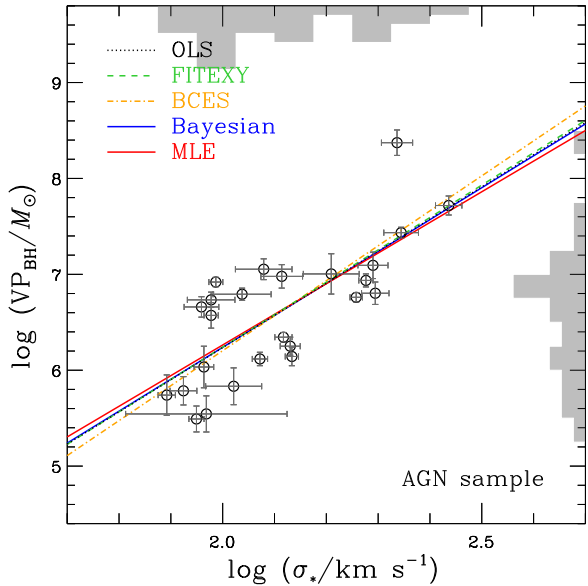}
\caption{
The $\rm VP_{\rm BH}-\sigma_*$ relation for the AGN sample (25).
As shown in the right-hand side grey histogram, we are currently suffering from a lack of high mass AGN sample.
Here we can clearly see the variability of the BCES estimator due to the effect of a single point (i.e., NGC 4253)
that is subject to much larger measurement error than others.
}
\label{msigma_agn}
\end{figure}

The virial factor is of fundamental importance for estimating AGN black hole masses  
in that it properly calibrates the measured virial product to a black hole mass
for both the reverberation mapping method and the single-epoch method.
Following Onken et al. (2004), we determine the average virial factor $\langle f \rangle$
by forcing the AGN black hole masses onto the \msigma~relation of quiescent galaxies.
%
%
%
The AGN sample used here is listed in Table~\ref{tab:AGNdata} with the corresponding references.
We updated the AGN sample given in Table 2 of Woo et al. (2010) by updating the virial products from 
Denney et al. (2010), revising the rms line widths from 
Park et al. (2012), and including the new $M_{\rm BH}$ estimate for Mrk 50 from Barth et al. (2011).
In Figure~\ref{msigma_agn}, we estimate the $\rm VP_{\rm BH}-\sigma_*$ relation with four regression methods, as in Figure~\ref{msigma_reproduce}. The regression results are listed in Table~\ref{tab:msigma-agn}.
The slope appears to be marginally lower than that for quiescent galaxies.
This small difference in slopes might be due to noise and AGN selection effects or it could be intrinsic, indicating a difference between passive and active galaxies (see Greene \& Ho 2006 and Schulze \& Wisotzki 2011).
Note that the current sample is not representative of the AGN population
since there is a deficit of high-mass AGNs, for which stellar velocity dispersion is extremely difficult to measure
due to the overwhelming AGN luminosity.

To determine the average virial factor we use the \texttt{FITEXY} estimator, fixing the intercept and
slope to be the same as that of the \msigma~relation for the quiescent galaxies,
\begin{equation}
{\chi ^2} = \sum\limits_{i = 1}^N {\frac{{{{\left( {{y_i} + \log f - \alpha  - \beta {x_i}} \right)}^2}}}{{\sigma _{y,i}^2 + {\beta ^2}\sigma _{x,i}^2 + \sigma _{{\mathop{\rm int}} }^2}}}.
\end{equation}
Here, $y = \log (\rm VP_{\rm BH} / M_{\odot})$ and $x = \log (\sigma_{\ast} / 200\ \mathrm{km~s^{-1}})$.
The free parameters are only $f$ and $\sigma _{{\mathop{\rm int}}}$.
Adopting the regression results listed in Table~\ref{tab:re-estimate}, we estimate the virial factor and list the
result in Table~\ref{tab:viral_factor}.

\begin{figure*} 
\centering
\includegraphics[width=15cm]{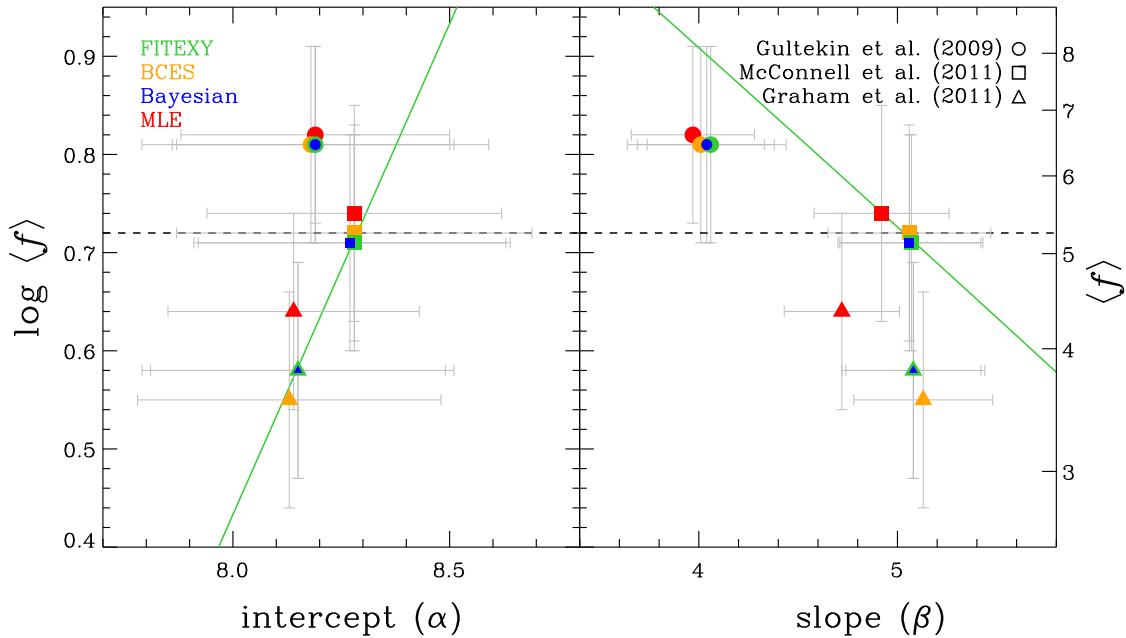} 
\caption{Variation of the estimated virial factor as a function of the adopted intercept (left) and slope (right) of 
the quiescent galaxy \msigma~relations taken from Table~\ref{tab:re-estimate}.
Symbols mean the corresponding dataset used for estimation of the intercept and slope 
as expressed in upper right corner, 
while the colors of symbols indicate the regression methods used for
them as given in the upper left corner.
The horizontal dashed line indicates the value of the virial factor estimated from Woo et al. (2010) (i.e., 0.72).
As an illustration, we add the green solid lines which show the dependence of the virial factor 
on the arbitrarily varied intercept (left) and slope (right) by fixing respectively the slope (left) 
and intercept (right) taken from the \texttt{FITEXY} estimates in the sample of McConnell et al. (2011).
}
\label{fig-f}
\end{figure*}

Figure~\ref{fig-f} shows the dependency of the virial factor on the adopted slope and intercept based on three datasets
with four regression methods.
As expected, the difference of $f$ between the different regression methods for a particular dataset is small, 
with the only exception being the value of $f$ obtained from MLE (red symbols).
Estimated virial factors vary as much as a factor of 2 among the data sets, 
larger than the typical uncertainties.
The difference in $f$ factors derived from the McConnell and Graham data is mostly due to the difference in the values from the sample of galaxies that overlap in these two data sets, as discussed in Section~\ref{sec:sample_issues}. 
The recent updates of \mbh~measurements by Schulze \& Gebhardt (2011) lead to a smaller mean mass in the Graham sample compared to that in the McConnell sample.
The difference of the adopted velocity dispersion measures 
results in a slightly larger velocity dispersion on average in the Graham sample compared to that of the McConnell sample.
These combined differences make the intercept of the \msigma~relationship smaller in the Graham sample than 
in the McConnell sample,
thus reducing the $f$ factor in the Graham sample regardless of the adopted regression methods.
As can be seen, the derived virial factor is susceptible to the small variation of the quiescent galaxy 
\msigma~relation within the current calibration process.

With the current AGN dataset, we are unable to constrain the $f$ factor
as a function of the mass range or host galaxy morphological type,
since the number of sources in our sample is small and the morphology
of our sample is biased toward late-type galaxies.
We note that a larger AGN sample (e.g., high-mass AGNs, especially) is needed for better statistical calibration.

\begin{deluxetable}{lccc}
\tablecolumns{4}
\tablewidth{182pt}
\tablecaption{The $\rm VP_{\rm BH}-\sigma_*$~Relation for the Active Galaxy Sample: 
$\log (\rm VP_{\rm BH} / M_{\odot}) = \alpha + \beta \log (\sigma_{\ast} / 200\ \mathrm{km~s^{-1}})$}
\tablehead{
\colhead{Method} & 
\colhead{$\alpha$}    & 
\colhead{$\beta$}       & 
\colhead{$\sigma_{{\mathop{\rm int}}}$} 
}
\startdata
\multicolumn{4}{c}{Forward regression} \\
\\
OLS       & $7.25\pm0.14$ & $3.35\pm0.57$ & \nodata \\
MLE       & $7.23\pm0.14$ & $3.20\pm0.59$ & $0.41\pm0.06$ \\
BCES      & $7.30\pm0.17$ & $3.65\pm0.75$ & $0.41\pm0.06$ \\
FITEXY    & $7.26\pm0.15$ & $3.38\pm0.63$ & $0.43\pm0.06$ \\
Bayesian  & $7.24\pm0.17$ & $3.33\pm0.69$ & $0.47\pm0.09$ \\
\hline
\multicolumn{4}{c}{Inverse regression} \\
\\
OLS       & $7.74\pm0.23$ & $5.93\pm0.82$ & \nodata \\
MLE       & $7.72\pm0.33$ & $5.88\pm1.21$ & $0.57\pm0.13$ \\
BCES      & $7.70\pm0.30$ & $5.72\pm1.10$ & $0.51\pm0.12$ \\
FITEXY    & $7.68\pm0.26$ & $5.68\pm0.95$ & $0.56\pm0.11$ \\
Bayesian  & $7.68\pm0.37$ & $5.67\pm1.87$ & $0.62\pm0.28$ 
\enddata
\label{tab:msigma-agn}
\end{deluxetable}

\begin{deluxetable}{lcc|cc}
\tablecolumns{5}
\tablewidth{230pt}
\tablecaption{The derived average virial factor and intrinsic scatter
based on the adopted \msigma~relation given in Table \ref{tab:re-estimate}}
\tablehead{
\colhead{Method} & 
\colhead{} &
\colhead{} &
\colhead{} &
\colhead{} \\
\colhead{} &
\colhead{$\log \langle f \rangle$}    & 
\colhead{$\sigma_{{\mathop{\rm int}}}$} &
\colhead{$\log \langle f \rangle$}    & 
\colhead{$\sigma_{{\mathop{\rm int}}}$} 
}
\startdata
& \multicolumn{4}{c}{From G\"ultekin et al. (2009) Sample} \\
& \multicolumn{2}{c}{using forward relation} & \multicolumn{2}{c}{using inverse relation} \\
\\
MLE       & $0.82\pm0.09$ & $0.43\pm0.05$ & $0.55\pm0.12$ & $0.54\pm0.06$  \\
BCES      & $0.81\pm0.10$ & $0.43\pm0.05$ & $0.60\pm0.11$ & $0.51\pm0.05$  \\
FITEXY    & $0.81\pm0.10$ & $0.43\pm0.05$ & $0.59\pm0.11$ & $0.52\pm0.06$  \\
Bayesian  & $0.81\pm0.10$ & $0.43\pm0.05$ & $0.57\pm0.12$ & $0.52\pm0.06$  \\
\hline
& \multicolumn{4}{c}{From McConnell et al. (2011) Sample} \\
& \multicolumn{2}{c}{using forward relation} & \multicolumn{2}{c}{using inverse relation} \\
\\
MLE       & $0.74\pm0.11$ & $0.48\pm0.05$ & $0.51\pm0.14$ & $0.62\pm0.07$  \\
BCES      & $0.72\pm0.11$ & $0.49\pm0.05$ & $0.52\pm0.13$ & $0.62\pm0.07$  \\
FITEXY    & $0.71\pm0.11$ & $0.49\pm0.05$ & $0.53\pm0.13$ & $0.61\pm0.06$  \\
Bayesian  & $0.71\pm0.11$ & $0.49\pm0.05$ & $0.52\pm0.13$ & $0.61\pm0.07$  \\
\hline
& \multicolumn{4}{c}{From Graham et al. (2011) Sample} \\
& \multicolumn{2}{c}{using forward relation} & \multicolumn{2}{c}{using inverse relation} \\
\\
MLE       & $0.64\pm0.10$ & $0.47\pm0.05$ & $0.42\pm0.13$ & $0.58\pm0.06$  \\
BCES      & $0.55\pm0.11$ & $0.50\pm0.05$ & $0.42\pm0.13$ & $0.57\pm0.06$  \\
FITEXY    & $0.58\pm0.11$ & $0.49\pm0.05$ & $0.45\pm0.12$ & $0.56\pm0.06$  \\
Bayesian  & $0.58\pm0.11$ & $0.49\pm0.05$ & $0.46\pm0.12$ & $0.56\pm0.06$ 
\enddata
\label{tab:viral_factor}
%
\end{deluxetable}

\section{Inverse fit} \label{sec:inverse}
In addition to the conventional forward fit relation (i.e., fitting $M_{\rm BH}$ for a given
$\sigma_*$, as we performed in previous sections), Graham et al. (2011) used an inverse fit for the 
\msigma~relation, suggesting that it corrects for possible sample selection bias due to non-detection
of intermediate-mass black holes ($<10^{6} M_{\odot}$).
We also follow their argument, refit all relations, and derive the average
virial factors based on the inverse relations (see Table~\ref{tab:re-estimate}, \ref{tab:msigma-agn}, and
\ref{tab:viral_factor}).
Note that basically forward and inverse fittings are not the same in the presence of intrinsic scatter.
 Depending on the direction of regression (i.e., whether to choose
 $M_{\rm BH}$ or $\sigma_*$ as the independent variable) the regressed slopes show 
large differences, leading to substantial changes in the virial factors. 
Therefore, we investigate and discuss the inverse fit in the context of the \msigma~relation.
Now we have three factors related to the linear regression, which make the problem more complicated:
measurements errors, intrinsic scatter, and truncation.

If there is a truncation in the y-axis (i.e., $\log M_{\rm BH}$) as argued by Graham et al. (2011), 
the conventional forward fit (fit y on x) causes a flattening in the
estimated slope due to the increased loss
of low mass black holes in the low $\sigma_*$ regime (e.g., see Appendix A in Mantz et al. 2010). 
The inverse fit (i.e., fit x on y) is not sensitive to this
Malmquist-type bias, so long as incompleteness only exist in black
hole mass.
As shown by Kelly (2007), in order to avoid this selection bias on the regression
result, it is necessary to assign the `independent variable' as the
variable used to select a sample.
This approach has been generally adopted in the Tully-Fisher relation studies since its sample is magnitude-selected
and errors are smaller in magnitude than in velocity
(e.g., Willick 1994; Tully \& Pierce 2000; Bamford et al. 2006; Weiner et al. 2006; Koen \& Lombard 2009; Williams et al. 2010; Miller et al. 2011).

In our sample, the measurement errors in the truncated coordinate ($M_{\rm BH}$) are larger
than in the other coordinate ($\sigma_*$).
Moreover, the sample selection might be highly inhomogeneous and simple selection
criteria may not be sufficient for describing it. The situation is more complex in the AGN sample selection. 
According to Schulze \& Wisotzki (2011), even though 
the inverse relation is insensitive to the mass-dependent selection, it does not yield the intrinsic
true relation without incorporating the knowledge of the underlying host galaxy distribution function,
which is currently hard to measure precisely. Furthermore, the AGN
sample likely exhibits incompleteness in $\sigma_*$ as well, as it is
harder to measure $\sigma_*$ for AGN hosting more massive
black holes due to their tendency to have higher luminosities. 
Thus there are good reasons to use either type of regression, 
but neither of them is completely free of bias.

We provide both regression results in Table~\ref{tab:re-estimate}, \ref{tab:msigma-agn}, and
\ref{tab:viral_factor}.
Inverse regression results in a steeper slope compared to that of forward regression in the \msigma~relation of 
quiescent galaxies.
The calibration based on the inverse regression makes black hole masses inferred from the AGN virial products
smaller, since most of the AGN sample is located at the low-mass regime,
thus leading to a reduction in the average virial factor.
This biased dependency toward the low-mass regime motivates expansion
in the dynamic range of sample of AGN containing both reverberation
mapping data and measurement of $\sigma_*$.
Based on these results, we conclude that the origin of the difference
in the recently reported virial factors (Woo et al. 2010 based on forward regression $vs.$ Graham et al. 2011 based on inverse regression) is mostly due to the direction of regression adopted 
(i.e., whether $M_{\rm BH}$ is considered the independent or dependent
variable), as well as the difference in the samples used to calibrate
the mass estimates.

Feigelson \& Babu (1992) suggested that we should choose the regression method for individual cases 
depending on the scientific question being investigated. Here the purpose of deriving the \msigma~relation 
for local quiescent galaxies is to calibrate AGN black hole masses with determining the virial factor.
By properly comparing the black hole
masses of inactive galaxies to virial products of active galaxies, the average virial factor is constrained as 
discussed in the previous section. Thus it is desirable to
  adopt the type of regression which yields the relation that
  minimizes the scatter in the black hole mass estimates (Graham \& Driver 2007).
It is more common to adopt the host spheroidal quantity as the
independent variable because the scaling relations are often used to
infer black hole mass using the host spheroidal quantities as a
proxy. Considering this, and the fact that the AGN sample likely
suffers from Malmquist bias in both $M_{\rm BH}$ and $\sigma_*$, we prefer
the calibration from the traditional forward regression.

\section{DISCUSSION AND CONCLUSIONS}\label{sec:conclusion}

We investigated the differences in the derived \msigma~relation and virial factor using the recently 
compiled datasets of quiescent and active galaxies.
The investigated possible origins of the difference are the fitting methodology and the sample difference.

For the difference in regression methods,
we utilized and compared four linear regression techniques: 
\texttt{FITEXY}, \texttt{Bayesian}, \texttt{BCES}, and \texttt{Maximum likelihood}.
With the current level of measurement errors of the \msigma~dataset, all estimators except for
the \texttt{maximum likelihood} estimator show good performance and consistent results with each other.
There is no significant difference between the estimators.
However, the assigned size of measurement errors on $\sigma_*$ can
have a significant impact on the regression results,
especially for the \texttt{BCES} and \texttt{Maximum likelihood} estimators.
The \texttt{Maximum likelihood} method using an implicit assumption of a uniform 
distribution for the intrinsic distribution of the independent
variables introduces a bias
which is clearly noticeable when the measurement errors on the independent variable are large
(e.g., above 10\% errors in the Graham sample as shown in Figure~\ref{fig-graph}).
Without properly accounting for the form of the intrinsic distribution of the independent variable,
MLE estimates are very similar to the OLS results.
Therefore we do not recommend this method for regression analysis in general.
Of course for the \msigma~regression analysis the difference in the estimated slope
is only marginal at the current adopted level of uncertainty on $\sigma_*$ ($5\%$).
The \texttt{BCES} estimator is also one of the good estimators based
on the current measurement error level on $\sigma_*$, 
although it may be problematic if the error is larger.
Based on our simulation results, the \texttt{FITEXY} estimator shows slightly better performance and
the least-biased result compared to the other methods, although the others also perform well and
the differences are marginal.
This is consistent with the result of Novak et al. (2006), although they did not provide an explicit implementation 
of all of the methods, nor a specific quantitative comparison.
In general, we recommend both the \texttt{FITEXY} and \texttt{Bayesian} estimators,
although the latter is computationally more intensive, especially when
the measurement errors are large.
However, we note that the \texttt{Bayesian} estimator has the
advantage over the method of \texttt{FITEXY} 
in that it calculates the full probability distribution function (i.e., posterior) of the parameters 
for the given data, and hence provides well-defined and reliable parameter uncertainties.
%
In addition, the \texttt{Bayesian} method can incorporate upper
limits, as can the method of G\"ultekin et al. (2009), 
whereas the \texttt{FITEXY} cannot.
If we perform the regression using the \texttt{Bayesian} method, 
for the G\"ultekin sample including upper limits as well as secure measurements, the result changes from 
($\alpha = 8.19\pm0.07$, $\beta = 4.04\pm0.40$, $\sigma_{\mathop{\rm int}} = 0.42\pm0.05$) to
($\alpha = 8.13\pm0.07$, $\beta = 4.32\pm0.38$, $\sigma_{\mathop{\rm int}} = 0.43\pm0.05$),
thus $\log \langle f \rangle$ correspondingly decreases from $0.81\pm0.10$ to $0.70\pm0.10$. 
As discussed in Tremaine et al. (2002), accurate and consistent estimation of an individual stellar velocity
dispersion with a correct measurement uncertainty is still required
and it will be an important factor
for better constraining the \msigma~relation and virial factor.

%
The difference in sample is the most important factor contributing to
the differences in derived \msigma~relations.
G\"ultekin et al. (2009) noted that the late-type galaxy and
pseudobulge population in their sample is the source of
the difference in intrinsic scatter measurements by comparing their sample to that of Tremaine et al. (2002).
Greene et al. (2010) found that the late-type low-mass galaxies show
large scatter and are offset relative to 
the \msigma~relation of elliptical galaxies using the sample of megamaser disk galaxies.
By extending the work of Graham et al. (2008), recently Graham et al. (2011) showed that the fraction of
barred galaxies in their sample alters the \msigma~relationship by dividing their sample into barred and non-barred galaxies.

\begin{figure} 
\centering 
\includegraphics[width=8cm]{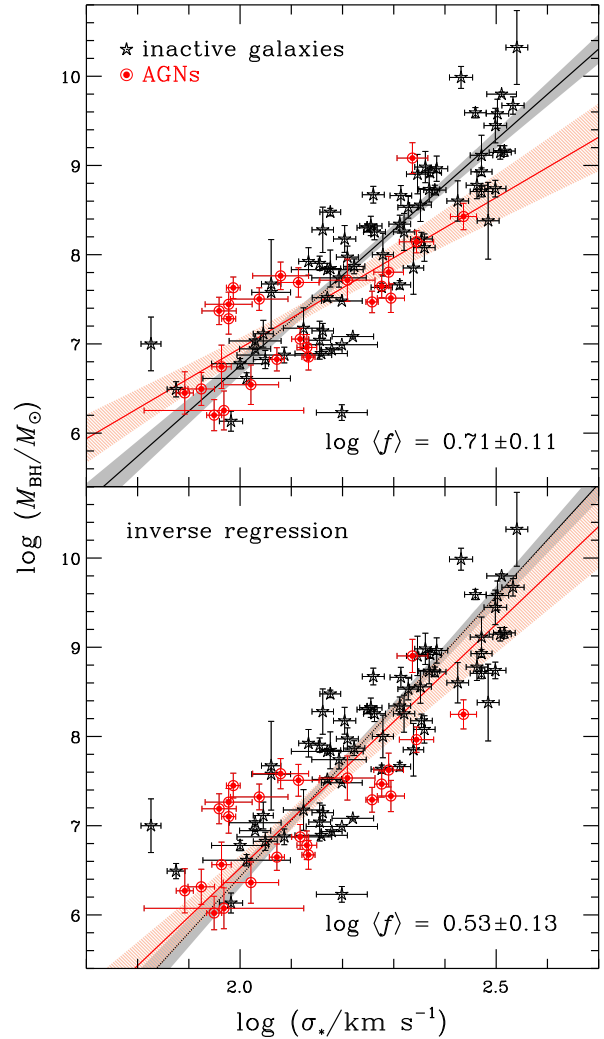}
\caption{
The updated $M_{\rm BH}-\sigma_*$ relations for the inactive (black) and active (red) samples using 
the \texttt{FITEXY} estimator for the forward regression (upper) and inverse regression (lower).
Shaded regions show the 1$\sigma$ (68\%) confidence intervals on the best-fit line.
The inactive sample is from McConnell et al. (2011) and is the most recent one. The active sample is given
in Table~\ref{tab:AGNdata}. 
}
\label{fig:final_msigma}
\end{figure}

According to these previous studies, the \msigma~relation seems to be not universal.
It varies depending on the mass range and galaxy type.
Correspondingly, the average $f$ factor is also significantly affected
by the sample population,
since the intercept and slope from the quiescent galaxy \msigma~relation are directly used in the calibration process.
As investigated in this study, the differences in the adopted sample contribute to the change of 
the virial factor.
Moreover, the direction of regression (forward {\it vs.} inverse) causes further changes in the virial factor. 
We showed that the derived $f$ factors vary as much as a factor of 2,
which is from a combined effect of the sample and regression
used. These differences could be thought of as an additional systematic uncertainty in the AGN black hole mass estimation
via the current calibration process of the virial factor,
since there is no obvious physical foundation for the selection of the
appropriate sample and direction of regression.

The true average $f$ factor should not be changed by the host galaxy type
since there should be no direct physical link between 
the AGN BLR geometry and the global morphology of host
galaxies. Unfortunately, the estimated average $f$ factor may be
subject to biases due to its calibration based on a single
\msigma~relationship. However, since the current sample is not large
enough to calibrate the virial factor as a function of galaxy type,
it is better to use a single value of the mean $f$ factor for AGN \mbh~estimation 
in order to avoid additional systematic errors 
until we get enough direct measurements of the structure of the BLR for 
an each individual AGN.
We note that an alternative method to measure AGN black hole mass that derives the virial factor 
through BLR modeling has been recently developed and applied to the reverberation data 
(e.g., Pancoast, Brewer, \& Treu 2011; Brewer et al. 2011; Pancoast et al. 2012).
Given the uncertainties in the $f$ factor, when investigating evolutionary trends in the \msigma~relation based on 
SE estimates, we recommend to use self consistent samples and techniques at different redshifts. 
In other words, one should measure the SE black hole masses consistently for AGN samples at different redshifts by using the cross-calibrated recipes based on the same virial factor. 
In this way the virial factor should be very similar for all samples and cancel out in the determination of the evolution of $\log M_{\rm BH}$ under the assumption that the virial factor is not a function of redshift (e.g., Woo et al. 2008).

Finally, we present the updated \msigma~relation for local active galaxies based on the \texttt{FITEXY}
estimator in Figure~\ref{fig:final_msigma}, where the forward (inverse) regression result is 
$\alpha = 7.97\pm0.14$, $\beta = 3.38\pm0.61$, and $\sigma_{\mathop{\rm int}} = 0.42\pm0.06$
($\alpha = 8.17\pm0.27$, $\beta = 5.47\pm1.01$, and $\sigma_{\mathop{\rm int}} = 0.52\pm0.11$).
The AGN black hole masses were converted from the virial products using the virial factor 
$\log \langle f \rangle = 0.71\pm0.11$ 
($\log \langle f \rangle = 0.53\pm0.13$)
derived in Section~\ref{sec:virial_factor}.
From a methodological point of view, we prefer the forward regression as discussed in Section~\ref{sec:inverse}. 
Thus our preferred value for the virial factor is 0.71 based on the preferred 
forward \texttt{FITEXY/Bayesian} estimation with the most recent sample (McConnell et al. 2011).
This value is consistent with that of Woo et al. (2010) (i.e., 0.72) and differs from that of Graham et al.
(2011) (i.e., 0.45) by $\sim 0.26$ dex. 
The difference arises from the combination of sample differences and regression differences.
It is worth noticing that the bottom panel of Figure~\ref{fig:final_msigma}
shows slightly better agreement between the non-AGN and AGN
\msigma~relations, which may indicate that the inverse regression has
less bias than the forward one and thus might be more
reliable. However, this conclusion only holds if we assume that the
active and inactive galaxies share the same
\msigma~relationship. Considering these issues, it is still not
conclusive whether the inverse method is preferable with the current
datasets owing to selection effects and limited dynamic range of the AGN sample.

\acknowledgments

This work was supported by the National Research Foundation of Korea (NRF) grant funded 
by the Korea government (MEST) (No. 2012-006087).
DP would like to thank Hyung Mok Lee, Matthew W. Auger, Jong Chul Lee, and Andreas Schulze for helpful discussions.
BK acknowledges support from the Southern California Center for Galaxy Evolution, a multi-campus research program funded by the University of California Office of Research.
TT acknowledges support from the Packard Foundation through a Packard research fellowship.
We thank the anonymous referee for constructive comments and suggestions.

\




\end{document}